\documentclass[runningheads]{llncs}
\usepackage{graphicx}
\usepackage[colorlinks=True,hypertexnames=false]{hyperref}
\usepackage{amsmath}
\usepackage{amssymb}
\usepackage{wasysym} 
\usepackage{soul} 
\newcommand{\mathcolorbox}[2]{\colorbox{#1}{$\displaystyle #2$}}
\usepackage{xcolor}
\usepackage{cancel} 
\usepackage{lscape}
\usepackage[left=0.7in,right=0.7in,top=1in,bottom=0.7in]{geometry} 
\usepackage{upgreek} 

\usepackage{nicematrix}

\begin{document}
\title{Intersectional inequalities in social networks}

\author{Samuel Martin-Gutierrez\inst{1} \and%
Mauritz N. Cartier van Dissel\inst{1,}\inst{2} \and%
Fariba Karimi\inst{2,}\inst{1}}
\authorrunning{S. Martin-Gutierrez}
%
\institute{Network Inequality Group, Complexity Science Hub, Josefstaedter Str. 39, 1080 Vienna, Austria
\and
Institute of Interactive Systems and Data Science, Graz University of Technology, Sandgasse 36, 8010 Graz, Austria}
\maketitle              
\begin{abstract}
Social networks are shaped by complex, intersecting identities that drive our connection preferences. These preferences weave networks where certain groups hold privileged positions, while others become marginalized. While previous research has examined the impact of single-dimensional identities on inequalities of social capital, social disparities accumulate nonlinearly, further harming individuals at the intersection of multiple disadvantaged groups. However, 
how multidimensional connection preferences affect network dynamics and in what forms they amplify or attenuate inequalities remains unclear.
In this work, we systematically analyze the impact of multidimensionality on social capital inequalities through the lens of intersectionality. To this end, we operationalize several notions of intersectional inequality in networks. Using a network model, we reveal how attribute correlation (or consolidation) combined with biased multidimensional preferences lead to the emergence of counterintuitive patterns of inequality that are unobservable in one-dimensional systems. We calibrate the model with real-world high school friendship data and derive analytical closed-form expressions for the predicted inequalities, finding that the model's predictions match the observed data with remarkable accuracy.
These findings hold significant implications for addressing social disparities and inform strategies for creating more equitable networks. 
\end{abstract}
%
%
%


\section{Introduction}

Our social ties are an invaluable resource that brings us a myriad of benefits. Close strong connections form support networks vital to maintaining our well-being \cite{putnamBowlingAloneCollapse2020}, while weak ties open doors to new opportunities and help us build our careers \cite{granovetterStrengthWeakTies1973,rajkumarCausalTestStrength2022}. Together, this collection of beneficial connections makes up our social capital \cite{colemanSocialCapitalCreation1988,linSocialCapitalTheory2001}. Our choice of ties depends heavily on our identity, as we tend to connect with people who are similar to us, following the homophily principle. Homophily is a very natural connection mechanism, as it leads to smoother coordination, better communication, and enhanced trust between individuals \cite{fuEvolutionHomophily2012}. However, recent studies have also shown that homophily can have pernicious effects, as it exacerbates segregation and creates inequalities in the social visibility of different groups \cite{jacksonInequalityEconomicSocial2021,karimiHomophilyInfluencesRanking2018,espin-noboaInequalityInequityNetworkbased2022}.
These inequalities of social capital then cause second-order disparities, as they affect income mobility \cite{chettySocialCapitalMeasurement2022}, educational disparities \cite{manzoEducationalChoicesSocial2013}, and health outcomes \cite{hiraokaHerdImmunityEpidemic2022}. Given the far-reaching implications of network effects on inequality \cite{zhaoNetworkDiffusionHomophily2021}, its study with explainable computational models has been identified as a research priority \cite{dimaggioNetworkEffectsSocial2012}.

So far, researchers have studied social network inequalities using simple network models with two interacting groups, a majority and a minority. They have found that inter-group inequality is mostly driven by group-based connection preferences, and in homophilic regimes (the most common in social networks), the majority group always has the advantage \cite{karimiHomophilyInfluencesRanking2018,espin-noboaInequalityInequityNetworkbased2022}. However, real-world systems can rarely be reduced to the interaction between two social groups. We are complex social beings with multidimensional identities, so our connections are also inherently multidimensional. The unresolved, yet crucial question is how do multidimensional interactions affect the emergence of multidimensional and intersectional inequalities of social capital in networks. Individuals can experience marginalization in several dimensions simultaneously (ethnicity, gender, socioeconomic status, etc.), and these disadvantages accumulate in a nonlinear way, further harming people at the intersection of several minority groups. Therefore, an intersectional lens is indispensable to understand these emergent inequalities \cite{crenshawDemarginalizingIntersectionRace1989}.
However, adopting an intersectional approach to measure social network inequalities presents significant challenges, as we still don't know in what mathematical forms inequalities are compounding and adding up. The research around intersectional inequality has been so far mostly conceptual and qualitative, with fewer efforts focused on developing quantitative methodologies \cite{bauerIntersectionalityQuantitativeResearch2021,zhaoSuperdiversityConsolidationImplications2023}.

In this work, we use synthetic and real-world multidimensional social networks to explore intersectional social capital inequalities. We use a multidimensional network model to generate synthetic networks and analyze high-school friendship datasets. To quantify inter-group inequalities, we adopt the number of neighbors of a node (the degree) as a proxy of social capital. We derive closed-form mathematical expressions for the model-predicted degree disparities, enabling a comprehensive understanding of degree inequalities.

The key insight of intersectionality is that advantages and disadvantages compound in complex ways \cite{IntersectionalityRevealingRealities}. At its heart, intersectionality rejects the idea that multidimensional inequalities are a simple additive composition of one-dimensional advantages or disadvantages \cite{hancockWhenMultiplicationDoesn2007}. As illustrated by Crenshaw in her seminal paper focusing on the experiences of black women, “Black women sometimes experience discrimination in ways similar to white women's experiences; sometimes they share very similar experiences with Black men. Yet often they experience double discrimination—the combined effects of practices which discriminate on the basis of race, and on the basis of sex. And sometimes, they experience discrimination as Black women—not the sum of race and sex discrimination, but as Black women” \cite{crenshawDemarginalizingIntersectionRace1989}.

Despite its importance, no systematic operationalization of intersectional inequalities in social networks has been proposed so far. To bridge this gap, we draw from the multifaceted definitions of intersectionality to develop operationalizations for three notions of intersectional inequalities. We focus on the idea that inequalities in multidimensional systems are fundamentally different and more complex than their one-dimensional counterparts. First, we consider \emph{simple intersectionality} as the quantitative differences between inequalities found in one- versus multidimensional systems. For example, we find that one-dimensional degree disparities take different values in multidimensional systems with respect to equivalent one-dimensional systems. Second, we define \emph{emergent intersectionality} as the new emergent properties of inequalities in multidimensional systems. For example, while in simple social systems with two interacting groups the minority is consistently disadvantaged in homophilic regimes and advantaged in heterophilic ones \cite{karimiHomophilyInfluencesRanking2018}, in certain multidimensional systems we counterintuitively find that minorities may be favored in both regimes, while majorities can be relegated to disadvantageous positions. Third, we develop a mathematical formulation to quantify the notion of inequality as the irreducibility of multidimensional inequalities to simple combinations of one-dimensional ones. We call these irreducible multidimensional disparities \emph{nonlinearly intersectional}, and we demonstrate that the degree disparities emerging from our simple network model have this property.

A crucial matter we need to consider when studying intersectional inequalities in social networks is attribute correlation. In his seminal work on intergroup relations, Blau discussed the profound effects of attribute correlation on connectivity patterns \cite{blauInequalityHeterogeneityPrimitive1977}, which in turn determine the positions and roles (advantaged or disadvantaged) of different groups in the network. For instance, if socioeconomic status (SES) and race are correlated, high homophily in SES will necessarily result in high rates of homophilous association in race, even if people's preferences were neutral in that dimension \cite{moodyRaceSchoolIntegration2001}, carrying over the associated inequalities across dimensions. Correlation thus heavily impacts intersectional inequalities, because preferences in a given dimension might spill over to another and indirectly generate disparities in an otherwise neutral dimension
\cite{garipNetworkAmplification2021,santiagoExtendedFormalismPreferential2008,leszczenskyEthnicSegregationFriendship2015,kronebergWhenEthnicityGender2021}. 
We control for attribute correlation by explicitly modeling the population distribution in our interaction model, which in turn controls for the pool of opportunities to meet people from different groups \cite{mcphersonBirdsFeatherHomophily2001,feldFocusedOrganizationSocial1981,sajjadiUnveilingHomophilyPool2024a}.

To conclude, we fit the network model to real-world social networks and compare the empirical values of inequality metrics to the theoretical ones, finding a remarkable alignment. The model can be used to project how the patterns of inequality would change if the characteristics of the system varied (population composition, connection preferences, etc.), enabling the accurate estimation of intersectional inequalities and the development of intervention scenarios \cite{neuhauserImprovingVisibilityMinorities2023}. 

\section{Network model of multidimensional group interactions}

Building on previous research on inequalities in social networks \cite{karimiHomophilyInfluencesRanking2018,espin-noboaInequalityInequityNetworkbased2022}, we use a network model to study how the disparities between social groups depend on their sizes and connection preferences. The key characteristic of the chosen network model that enables the study of intersectional inequalities is that it provides a transparent and tractable representation of multidimensional interactions \cite{martin-gutierrezHiddenArchitectureConnections2024}.

Let us consider a directed network where each node $i$ belongs to a social group defined by a multidimensional vector of attributes $\mathbf{s}\in \mathcal{S}=S_1\times S_2\times \dots \times S_D=\{1,\dots,v_1\}\times \{1,\dots,v_2\}\times \dots \times \{1,\dots,v_D\}$, so that we have $D$ dimensions and $s^d \in \{1,\dots,v_d\}$ is the value of $i$'s attribute in dimension $d$, which has $v_d$ different possible values. Therefore, there are $n_{mg}=\prod_{d=1}^{D} v_d$ multidimensional groups. 
{For example, in a context where sex and nationality are the relevant dimensions, with sex having two possible values (female - $\female$ and male - $\mars$) and nationality three ((A)rmenian, (B)razilian, and (C)hinese), the possible attribute vectors describing multidimensional groups would be $\mathbf{s} = \{(\female, A), (\female, B),(\female, C), (\male,A), (\male, B), (\male, C) \}$.}
We encode the population distribution in a tensor $F$ where the element $F_{s^1,s^2,\dots,s^D}$ is the fraction of individuals in the multidimensional group $\mathbf{s}=(s^1, s^2\dots,s^D)$ and $\sum_{s^1,s^2,\dots,s^D} F_{s^1,s^2,\dots,s^D} = 1$.

Every node also has some latent connection preferences associated with each dimension. We encode these connection preferences in $D$ homophily/hetereophily matrices $h^d$, one for each of the $D$ dimensions. In the example above, we would have a $2\times2$ matrix for sex and a $3 \times 3$ matrix for nationality. Together, the population fractions tensor $F$ and the $D$ preference matrices $h^d$ constitute the parameters of the model.

\begin{figure}[h!]
\centering
\includegraphics[width=1.0\textwidth]{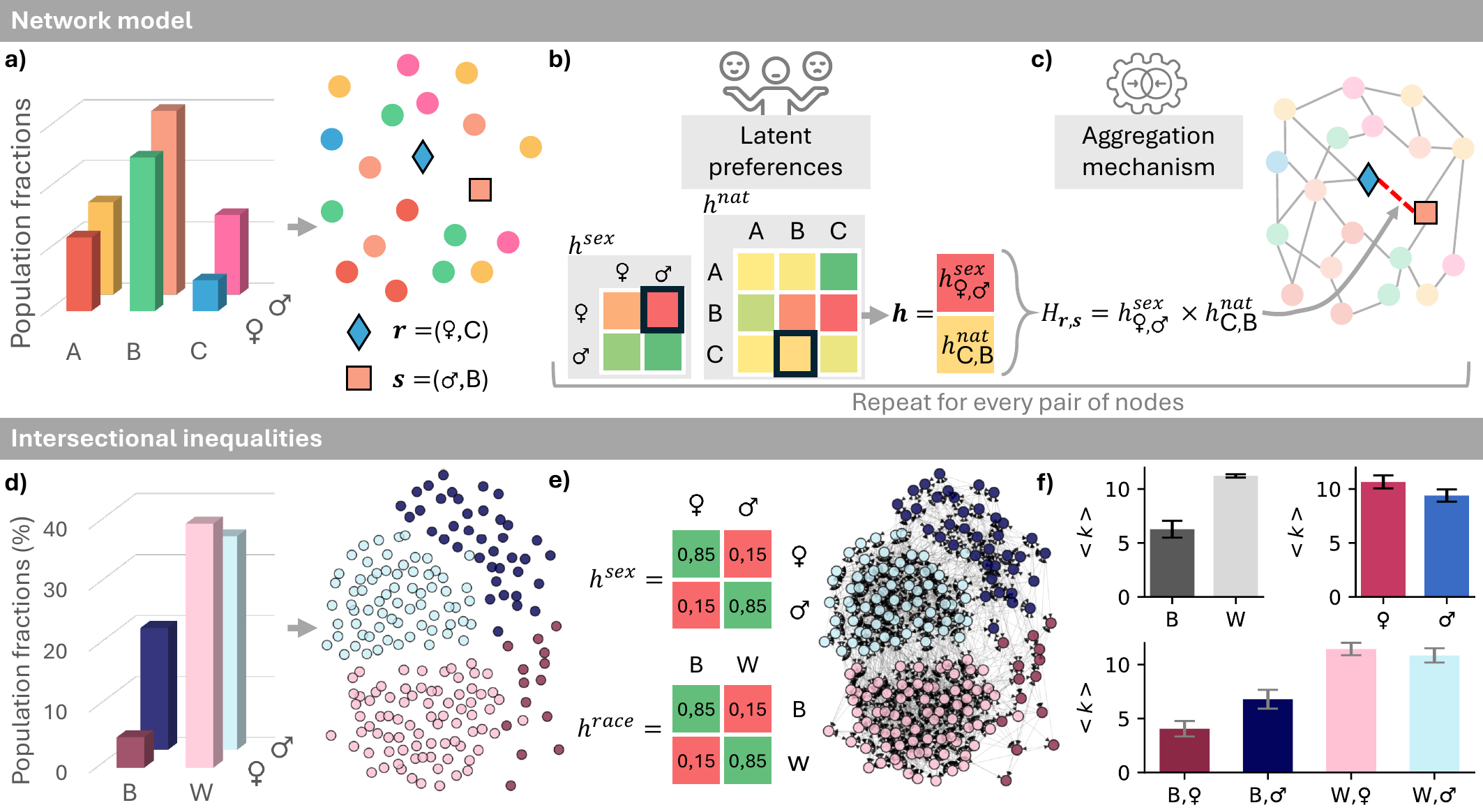}
\caption{\textbf{The multidimensional network model and emerging intersectional inequalities}. Panels \textbf{a}-\textbf{c} show the network formation dynamics in a two-dimensional system where the relevant dimensions are sex (female $\female$ or male $\mars$) and nationality ((A)rmenian, (B)razilian, (C)hinese). In panel \textbf{a}, we generate a set of $N$ nodes sampling from the population distribution $F$ and pick a pair of nodes belonging to group $\mathbf{r}$hombuses and group $\mathbf{s}$quares. In panel \textbf{b}, we build group $\mathbf{r}$'s latent preference vector $\mathbf{h}$ for group $\mathbf{s}$. In panel \textbf{c}, the preferences are combined through the aggregation mechanism of Eq. \eqref{eq:all_dim}, leading to a tie formation probability $H_{\mathbf{r},\mathbf{s}}$ that determines whether a link is established between the two nodes. Panels \textbf{d}-\textbf{f} show a synthetic two-dimensional network and the average degree $\left< k\right>$ inequalities that emerge in such system. The two dimensions are race ((B)lack or (W)hite) and sex ($\female$ or $\male$). Panel \textbf{d} is analogous to panel  \textbf{a}. Panel \textbf{e} shows the latent preference matrices $h$ and a representative network where we fixed the average degree of each node to 10. In panel \textbf{f}, we compute the average degree for each one-dimensional and multidimensional group, finding that black females suffer an intersectional disadvantage, as their degree is substantially lower than the average degree of black people and females. We have averaged the results from 100 simulations. The error bars show the standard error of the mean.}
\label{fig:model}
\end{figure}

To generate a network, we consider a fixed set of $N$ nodes with attribute vectors picked from the population distribution $F$,  as illustrated in panel \textbf{a} of Fig.~\ref{fig:model}. Then, we pick a random pair of nodes $i$ and $j$. Let us assume that node $i$ belongs to multidimensional group $\mathbf{r}$  (we can call it rhombuses) and node $j$ belongs to multidimensional group $\mathbf{s}$ (squares). Thus, the latent preferences of $i$ would be $\{h^d_{r^d,s^d};d=1, 2, \dots,D\}$ (see Fig.~\ref{fig:model}\textbf{b}). We treat the latent preferences in the $h^d_{r^d,s^d}$ matrices as \emph{partial probabilities} of establishing a link. When two nodes $i$ and $j$ meet, node $i$ evaluates its preference in each dimension $h^d_{r^d,s^d}$ independently (it \emph{tosses a biased coin} with success probability $h^d_{r^d,s^d}$), and decides to make a link to $j$ if all the independent evaluations are successful. Since the evaluations are performed independently for each dimension, the final probability of making a link is: 

\begin{equation}
    H_{\mathbf{r},\mathbf{s}}  = \prod_{d=1}^D h^d_{r^d,s^d}
    \label{eq:all_dim}
\end{equation}

For instance, in the example above, the connection probability of a node of group $\mathbf{r} = (\female,C)$ to connect to a node of group $\mathbf{s} = (\male,B)$ would be $H_{(\female,C), (\male,B)} = h^{\text{sex}}_{\female,\male} h^{\text{nat}}_{C,B}$, as shown in Fig.~\ref{fig:model}\textbf{c}.
We make a tie-formation attempt for each of the $N(N-1)$ possible node pairs. The plausibility of this tie-formation mechanism as an accurate description of multidimensional social dynamics has been validated with real-world networks \cite{martin-gutierrezHiddenArchitectureConnections2024}.

In this directed network, an edge $i\rightarrow j$ represents node $i$'s choice to connect to node $j$. We use the in-degree $k_j$ (the number of incoming connections to node $j$) as a proxy for $j$'s social capital, as it captures aspects of popularity, visibility, and access to social support. In panels \textbf{d}-\textbf{f} of Fig.~\ref{fig:model} we show the intersectional degree inequalities that emerge in a synthetic two-dimensional system where the dimensions are race (Black or White) and sex ($\female$ or $\male$), with Black people and women being the minorities in each dimension. As shown in Fig.~\ref{fig:model}\textbf{f}, Black women, while already being at a disadvantage due to being Black, are further marginalized, being relegated to the periphery of the network and gathering almost half as many links as Black men and almost a third compared to White people. This is particularly surprising considering that women are slightly advantaged over men in this system. Disentangling the intricacies of such intersectional inequality patterns is one of our central objectives.

\section{Modeling consolidation and group size distribution}

A key insight of previous studies is that inter-group network inequalities are caused by the combination of two mechanisms: imbalanced group sizes and biased connection preferences \cite{karimiHomophilyInfluencesRanking2018}. As argued by Feld \cite{feldFocusedOrganizationSocial1981}, the attribute distribution among the pool of people available for forming links can solely determine the structure of social networks, potentially leading to the emergence of homogeneous networks with many in-group links and few out-group links even when preferences are neutral.
Moreover, as Blau discusses in his seminal works on group interactions \cite{blauInequalityHeterogeneityPrimitive1977}, network segregation can be further exacerbated if attributes are correlated.
This highlights the importance of considering the effect of the population distribution when studying intersectional inequalities in social networks. In combination with biased connection preferences, attribute correlation can have a dramatic impact on the visibility of minorities.

In this section, we develop a systematic method to generate population distributions with specific group sizes and build a taxonomy of distributions for two-dimensional binary systems ($D=2, v_1=v_2=2$).
Let us start with the constraints imposed by the one-dimensional marginal distributions, which will determine the possible sizes of multidimensional groups (whether they are a minority or a majority in the system). Without loss of generality, within each dimension one group will be the minority $m$ and the other the majority $M$, so we have multidimensional groups $(m,m),(m,M),(M,m),$ and $(M,M)$. The population tensor (a matrix in this case) takes the following form:

\begin{equation}
    \begin{NiceMatrixBlock}[auto-columns-width]
    F = 
        \begin{bNiceMatrix}
            F_{m,m} & F_{m,M} \\
            F_{M,m} & F_{M, M}
        \end{bNiceMatrix}
    \end{NiceMatrixBlock}
\end{equation}

Where $F_{m,M}$ is the fraction of individuals belonging to group $(m,M)$. We define the marginal distribution for a one-dimensional group $s^1$ in dimension $1$ and a group $s^2$ in dimension $2$ as

\[ f_{s^1}^1 = F_{s^1,m} + F_{s^1,M} \qquad \qquad f_{s^2}^2 = F_{m,s^2} + F_{M,s^2} \]

With $s^1, s^2 \in \{m,M\}$. For instance, if the population tensor is:

\begin{equation}
    \begin{NiceMatrixBlock}[auto-columns-width]
    F = 
        \begin{bNiceMatrix}
            0.15 & 0.20 \\
            0.30 & 0.35
        \end{bNiceMatrix}
    \end{NiceMatrixBlock}
\end{equation}

The one-dimensional population fractions would be $\{f^1_m=0.35, f^1_M=0.65\}$ and $\{f^2_m=0.45, f^2_M=0.55\}$. By definition, the minority in each dimension has size $f^d_m \leq 0.5$, and the majority, $f^d_M \geq 0.5$.  Let us further assume that $f^1_m \leq f^2_m$ (the minority of the first dimension is the smallest group) and notice that $f^d_M = 1 - f^d_m$. We can always organize one-dimensional groups in this way by (re)labeling them adequately.

With these constraints, the population tensor can take two extreme forms
(and many between them)
, one with the highest values in the diagonal, which we call distribution of maximum correlation ($MC$), and another with the highest values in the anti-diagonal, called distribution of maximum anti-correlation ($AC$): 

\begin{equation}
    \begin{NiceMatrixBlock}[auto-columns-width]
    F^{MC} =  
        \begin{bNiceMatrix}
        f^1_m & \quad 0 \\
        f^2_m-f^1_m & \quad 1-f^2_m
        \end{bNiceMatrix}
    \qquad
    F^{AC} = 
        \begin{bNiceMatrix}
        0 &\quad f^1_m \\
        f^2_m &\quad     1-f^2_m-f^1_m
        \end{bNiceMatrix}
        \end{NiceMatrixBlock}
\end{equation}


The constraints on the marginal population distributions ($f^1_m \leq f^2_m \leq f^2_M \leq f^1_M$) also impose constraints on the relative sizes of the multidimensional groups:

\begin{align} 
F^{MC}_{M,M} \geq F^{MC}_{M,m}, F^{MC}_{m,m} \geq F^{MC}_{m,M}\\
F^{AC}_{M,m} \geq F^{AC}_{m,M}  \geq F^{AC}_{m,m} \\
F^{AC}_{M,M} \geq F^{AC}_{m,m}
\end{align}

Depending on the relative minorities' sizes $f^1_m$ and $f^2_m$,  we have the following additional set of inequalities:

\begin{alignat}{2} 
\label{eq:first_constraint}
\text{If} & \quad f^2_m >  2 f^1_m & \implies & F^{MC}_{M,m} > F^{MC}_{m,m} \\
\label{eq:second_constraint}
\text{If} & \quad f^2_m <  \frac{1}{2}(1-f^1_m) & \implies & F^{AC}_{M,M} > F^{AC}_{M,m} \\
\text{If} & \quad \frac{1}{2} (1-f^1_m) < f^2_m <  1 - 2f^1_m & \implies & F^{AC}_{m,M} < F^{AC}_{M,M} < F^{AC}_{M,m}
\label{eq:last_constraint}
\end{alignat}

\begin{figure}[h!]
\centering
\includegraphics[width=1.0\textwidth]{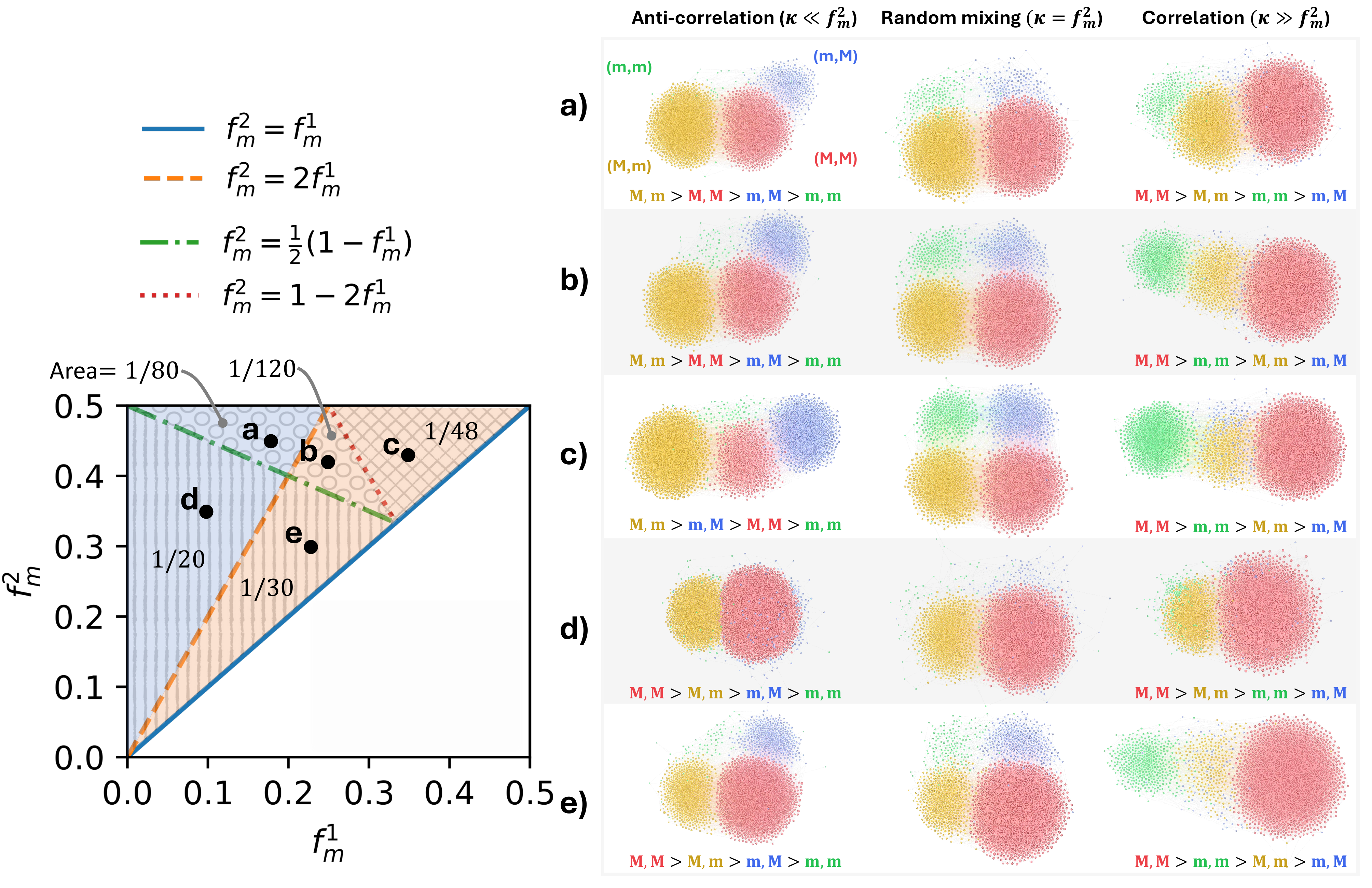}
\caption{\textbf{Effect of the population distribution on the network structure}. Two-dimensional system with two attributes per dimension ($m$ and $M$), and one-dimensional preferences defined in Eq. \eqref{eq:1dpref_example}. The left panel shows the regions of the $(f^1_m, f^2_m)$ space defined by the inequalities of Eqs. \eqref{eq:first_constraint}-\eqref{eq:last_constraint}, which result in the size rankings shown below the networks of the right. The plots of the right-hand side are representative examples of networks generated using the parameters from the \textbf{a}-\textbf{d} points of the left panel, plotted using a force-directed algorithm.}
\label{fig:effect_of_population_distribution}
\end{figure}

These inequalities effectively divide the two-dimensional space $(f^1_m, f^2_m) \in [0,0.5] \times [0,0.5]$ in the regions shown in the left-hand side of Fig.~\ref{fig:effect_of_population_distribution}, where colors mark regions divided by Eq. \eqref{eq:first_constraint} and patterns mark regions divided by Eqs. \eqref{eq:second_constraint}-\eqref{eq:last_constraint}.
Given the disparate extension of the regions, some orderings or size rankings of the multidimensional groups may be more common than others; for example, situations where $(M,M)$ is not among the two largest groups may be rare, as this only happens in region \textbf{c} with high anti-correlation. A comprehensive mathematical characterization of these bivariate distributions can be found in \cite{nguyenGeometryCertainFixed1985}.

Although the marginals constrain the possible values of the population distribution, they do not fully determine it. In the case of our 2D system with $v_d=2$ values per dimension, we need another parameter, which controls the correlation between both dimensions. Thus, we introduce a correlation parameter $\kappa \in  [0,1]$ that interpolates between the two extreme distributions:

\begin{equation}
    F = \kappa F^{MC} + (1-\kappa) F^{AC} = 
        \begin{bNiceMatrix}
        \kappa f^1_m &\quad (1-\kappa) f^1_m \\
        f^2_m - \kappa f^1_m &\quad     1-f^2_m-(1-\kappa)f^1_m
        \end{bNiceMatrix}
\end{equation}

This correlation parameter is equivalent to the concept of consolidation. In a population with uncorrelated attributes, each element of $F$ is the product of the corresponding marginals $F_{s^1,s^2} = f^1_{s^1} f^2_{s^2}$; in particular, $F_{m,m}=f^1_m f^2_m$. Therefore, the random mixing case corresponds to $\kappa = f^2_m = \max(f^1_m, f^2_m)$. Notice that for a general system with $v_1,v_2,\dots,v_D$ values per dimension, we need $\prod_d v_d - \left[ \sum_d v_d-(D-1) \right]$ parameters in addition to the marginal distributions to fully define the population distribution, so we can only tune consolidation with a single parameter in the case where $D=2; v_1=v_2=2$.



In Fig.~\ref{fig:effect_of_population_distribution}, we illustrate the combined impact of the minority sizes $f^1_m, f^2_m$, and attribute correlation on synthetic networks of a two-dimensional system with two attributes per dimension and the follwoing one-dimensional preference matrices:

\begin{equation}
    \begin{NiceMatrixBlock}[auto-columns-width]
    {h^1_{r^1,s^1}} = {h^2_{r^2,s^2}}= 
        \begin{bNiceMatrix}
            0.85 & 0.15 \\
           0.15 & 0.85
        \end{bNiceMatrix}
    \end{NiceMatrixBlock}
    \label{eq:1dpref_example}
\end{equation}

Even if the connection preferences do not change, their interaction with the relative sizes of the groups leads to vastly different network topologies. When there is no correlation (central panels of Fig.~\ref{fig:effect_of_population_distribution}), there are four independent groups, with the smaller groups acting as a periphery (panels \textbf{a}, \textbf{d}, and \textbf{e}). When there is high correlation (right-hand side panels), the groups $(m,m)$ and $(M,M)$ polarize the network if they are big enough, with the other groups acting as bridges.  We observe a similar behavior when there is high anti-correlation (left-hand side panels), with groups $(M,m)$ and $(m,M)$ being at two extremes and the other two acting as bridges in panels \textbf{a}, \textbf{b}, and \textbf{c}. When the sizes of $(m,M)$ and $(m,m)$ are too small, the network structure is fully dominated by the other two groups, as in panels \textbf{d} and \textbf{e}. This brief exploration of the diverse network structures that can emerge in a simple two-dimensional system shows how attribute correlation impacts network integration, as anticipated by Blau \cite{blauInequalityHeterogeneityPrimitive1977} and more recently by Garip and Molina \cite{garipNetworkAmplification2021}.

Correlation sometimes has counterintuitive effects, such as when the multidimensional majority group $(M,M)$ is the second smallest (panel \textbf{c}, left), or when the multidimensional minority group $(m,m)$ becomes the second largest (panels \textbf{b}, \textbf{c}, \textbf{e}, right). Another less striking but crucial observation is that simultaneously belonging to the smallest minority and the smallest majority $(m,M)$ puts individuals at a size disadvantage in most situations. In fact, if we sort the four multidimensional groups by size and sum the (reversed) ranks of the groups in the maximum correlation and anti-correlation situations, with rank $0$ being assigned to the smallest and rank $3$ to the largest (e.g. in panel \textbf{a} left, we have $(M,m):3,(M,M):2,(m,M):1,(m,m):0$), we find that the aggregate rank sum of group $(m,M)$ for these 10 scenarios is the lowest:  $\text{RS}_{m,M}=6 < \text{RS}_{m,m}=8 < \text{RS}_{M,m}=20 < \text{RS}_{M,M}=26$.
If we further weight the ranks by the area covered by each region (the triangles marked with a combination of color and pattern in Fig. \ref{fig:effect_of_population_distribution} left), we find $\text{RS}^w_{m,M}=7/48 < \text{RS}^w_{m,m}=9/48 < \text{RS}^w_{M,m}=23/48 < \text{RS}^w_{M,M}=33/48$.

\section{Intersectional inequalities in multidimensional social networks}

From the analyses performed in the previous section, it is clear that the visibility of a group within a network depends not only on the connection preferences but also on group sizes and correlation. 
The combination of biased connection preferences and imbalanced group sizes lead to remarkably complex inequality patterns in multidimensional systems (compared to previously studied one-dimensional systems \cite{karimiHomophilyInfluencesRanking2018,espin-noboaInequalityInequityNetworkbased2022}), with different groups performing diverse roles and occupying advantaged or disadvantaged positions within the network.



This sort of complexity was exposed by Crenshaw in her groundbreaking study on inequalities at the intersection of sex and race \cite{crenshawDemarginalizingIntersectionRace1989,crenshawMappingMarginsIntersectionality1991}. She introduced the concept of intersectionality to highlight the fundamental differences between the inequalities that emerge in one-dimensional versus multidimensional societies. Intersectionality emphasizes that often multidimensional inequalities cannot be reduced to a mere combination of one-dimensional ones. This observation has important consequences in how we should approach fairness in multidimensional social and algorithmic systems. 

Recognizing this complexity, we explore multidimensional inequalities in social networks through an intersectional lens. However, given the breadth of the concept of intersectionality, rather than using a single metric to characterize it, we propose several operationalizations corresponding to different aspects:

\begin{itemize}
    \item \emph{Simple intersectionality}: how the magnitude of inequalities in multidimensional systems differs from comparable one-dimensional systems, even when measuring the same type of disparity.
    \item \emph{Emergent intersectionality}: how multidimensional systems give rise to entirely new patterns of inequality that cannot exist in one-dimensional systems. 
    \item \emph{Nonlinear intersectionality}: how inequalities in multidimensional systems cannot be predicted by simply combining the inequalities observed in each dimension separately. 
\end{itemize}


\subsection{Measuring social capital inequalities}

We characterize social capital inequalities by measuring inter-group in-degree disparities. To this end, we use the \emph{stochastic difference} between the groups' degree distributions, a metric based on the probability for the in-degree of a random node $i$ of group $R (Rhombuses)$, $k_{i\in R}$, to be higher (or lower) than the in-degree of a random node $j$ from another group $S (Squares)$, $k_{j \in S}$ 
\cite{varghaCritiqueImprovementCL2000}:

\begin{equation}
    \delta_{R,S} = P(k_{i \in R} > k_{j \in S}) - P(k_{i \in R} < k_{j \in S})
\label{eq:inequality_metric_pairwise}
\end{equation}

We further define a \emph{total disparity} metric to quantify the overall (dis)advantage of a given group with respect to the rest as follows:

\begin{equation}
    \delta_R = P(k_{i \in R} > k_{j \notin R}) - P(k_{i \in R} < k_{j \notin R}) 
\label{eq:inequality_metric}
\end{equation}

Here, we are using the group names $R, S$ to represent the set of nodes belonging to them. $R$ and $S$ can be one-dimensional groups, multidimensional groups, or any other pair of disjoint sets of nodes. The metric $\delta_R \in [-1,+1]$ naturally induces a hierarchy in the groups, as the higher $\delta_R$, the more privileged group $R$ is. While $\delta_R$ is a group-level metric, it captures the (dis)advantage experienced by a (random) individual node in group $R$.

In Methods, we derive a closed-form analytical expression for the advantage of a multidimensional group $\mathbf{r}$ over another group $\mathbf{s}$ using a Poissonian approximation:

\begin{equation}
    \delta_{\mathbf{r}, \mathbf{s}} = Q_1(\sqrt{2 \lambda_{\mathbf{r}} }, \sqrt {2 \lambda_{\mathbf{s}}}) - Q_1(\sqrt{2 \lambda_{\mathbf{s}} }, \sqrt {2 \lambda_{\mathbf{r}}})
\label{eq:pairwise_multi_ineq}
\end{equation}

Where $Q_\nu(a,b)$ is the generalized Marcum Q-function of order $\nu$ and $\lambda_{\mathbf{r}}$ is the expected in-degree of group $\mathbf{r}$:

\begin{equation}
    \lambda_\mathbf{r} = N \sum_{\mathbf{s}} F_\mathbf{s} H_{\mathbf{s},\mathbf{r}}
\label{eq:expected_indegree}
\end{equation}

Notice that in-degree inequalities not only depend on preferences ($H_{\mathbf{r},\mathbf{s}}$) and population fractions ($F_\mathbf{s}$), but also on network size ($N$). 
Taking advantage of the properties of $\delta$, we have derived closed-form expressions for the pairwise disparities between two one-dimensional groups ($\delta^d_{r^d, s^d}$) and the total disparity of multidimensional ($\delta_{\mathbf{r}}$) and one-dimensional groups ($\delta^d_{r^d}$):


\begin{align}
\label{eq:pairwise_onedim_ineq}
\delta^d_{r^d, s^d} = &%
    \frac{1}{f^d_{r^d} f^d_{s^d}}%
    \sum_{\substack{\rho^d=r^d,\\\sigma^d=s^d}}
    F_{\boldsymbol{\rho}} F_{\boldsymbol{\sigma}} \delta_{\boldsymbol{\rho}, \boldsymbol{\sigma}} \\
    \label{eq:total_multi_ineq}
    \delta_\mathbf{r} = & \frac{1}{1- F_\mathbf{r}} \sum_{\boldsymbol{\sigma} \neq \mathbf{r}} F_{\boldsymbol{\sigma}} \delta_{\mathbf{r}, \boldsymbol{\sigma}} \\%
    \label{eq:total_onedim_ineq}
    \delta^d_{r^d} = &
    \frac{1}{f^d_{r^d} (1-f^d_{r^d})}%
    \sum_{\substack{\rho^d=r^d,\\\sigma^d \neq r^d}} F_{\boldsymbol{\rho}} F_{\boldsymbol{\sigma}} \delta_{\boldsymbol{\rho}, \boldsymbol{\sigma}}
\end{align}

In Eq. \eqref{eq:pairwise_onedim_ineq}, the sum $\sum_{\rho^d=r^d, \sigma^d=s^d}$ is performed over all the $\boldsymbol{\rho}$ and $\boldsymbol{\sigma}$ vectors whose component $d$ are respectively $r^d$ and $s^d$. In Eq. \eqref{eq:total_multi_ineq}, the sum runs over all $\boldsymbol{\sigma}$ vectors with any component different from $\mathbf{r}$. In Eq. \eqref{eq:total_onedim_ineq}, we sum over all $\boldsymbol{\rho}$ vectors whose $d$ component is $r^d$ and all $\boldsymbol{\sigma}$ vectors whose $d$ component is different from $r^d$.

These expressions allow us to directly calculate inter-group degree disparities from the model parameters (the population distribution tensor $F$ and the latent preference matrices $h^d$), eliminating the need for network simulations.
We just need to obtain the multidimensional preference matrix $H_{\mathbf{r},\mathbf{s}}$ from the latent preference matrices $h^d$ using Eq. \eqref{eq:all_dim}, plug it into Eq. \eqref{eq:expected_indegree}, and compute all pair-wise degree disparities with Eq. \eqref{eq:pairwise_multi_ineq}, which we can finally use to find total multidimensional disparities from Eq. \eqref{eq:total_multi_ineq} and one-dimensional disparities from Eq. \eqref{eq:total_onedim_ineq}. We provide comprehensive formulations of Eqs. \eqref{eq:pairwise_onedim_ineq} - \eqref{eq:total_onedim_ineq} expressed as a function of the population distribution $F_\mathbf{s}$ and latent preference matrices $h^d_{r^d,s^d}$ in Eqs. \eqref{eq:pairwise_onedim_ineq_full}-\eqref{eq:total_onedime_ineq_full} in Methods. These analytical expressions are in excellent agreement with network simulations, as we show in Fig.~\ref{fig:examples_2D_inequality}.

\subsection{Intersectional inequalities in a two-dimensional system}

To explore multidimensional inequalities, we again consider a two-dimensional system with two categories per dimension ($D=2,v_1=v_2=2$). Following the same terminology as before, we label the one-dimensional groups as $m$ (minority) or $M$ (Majority), and we always assign the smaller minority to the first dimension. Even in the simplest two-dimensional system, we have a plethora of parameter combinations that generate a wide variety of scenarios. To narrow down our analyses and illustrate the most significant phenomena, we discuss a selection of scenarios with different population distributions $F$ and the following symmetrical one-dimensional preference matrices, where we tune the $\mathsf{h}$ parameter to transition between a heterophilic ($\mathsf{h}=0$) and a homophilic ($\mathsf{h}=1$) regime:

\begin{equation}
    \begin{NiceMatrixBlock}[auto-columns-width]
    h^1_{r^1,s^1} = h^2_{r^2,s^2}= 
        \begin{bNiceMatrix}
            \mathsf{h} & 1-\mathsf{h} \\
           1-\mathsf{h} & \mathsf{h}
        \end{bNiceMatrix}
    \end{NiceMatrixBlock}
    \label{eq:1dpref_tune_h}
\end{equation}

\begin{figure}[h!]
\centering
\includegraphics[width=1.0\textwidth]{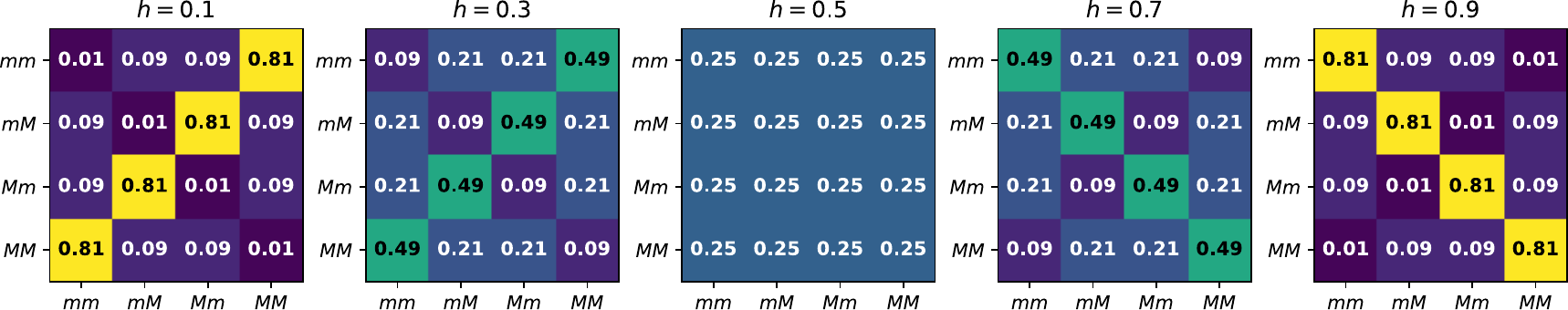}
\caption{\textbf{Examples of multidimensional preference matrices}. $H_{\mathbf{r},\mathbf{s}}$ matrices computed from the one-dimensional preference matrices of Eq. \eqref{eq:1dpref_tune_h} and the aggregation function of Eq. \eqref{eq:all_dim}.}
\label{fig:H_matrix}
\end{figure}

Fig.~\ref{fig:H_matrix} shows the resulting multidimensional preference matrices $H_{\mathbf{r},\mathbf{s}}$ for different values of the $\mathsf{h}$ parameter. 

Adopting an intersectional perspective, we will explore the most unexpected inequality patterns considering our current knowledge of inequalities in one-dimensional systems with two interacting groups. According to the literature \cite{karimiHomophilyInfluencesRanking2018,espin-noboaInequalityInequityNetworkbased2022}, larger groups typically have an advantage over smaller ones in homophilic regimes while smaller groups have the advantage in heterophilic regimes. Despite the complexity of multidimensional systems, we can understand the observed inequalities by considering the following general principles:

\begin{enumerate}
    \item Other things being equal, group $R$ has an advantage over group $S$ when groups have a higher preference to connect to $R$ than to $S$. 
    \item Other things being equal, group $R$ has an advantage over group $S$ if groups connecting to $R$ with high preference are bigger than groups connecting to $S$ with high preference. 

\end{enumerate}

Although these principles do not explain the totality of situations arising in multidimensional systems, they help us understand most of them. For example, all the inequality patterns found in previous studies \cite{karimiHomophilyInfluencesRanking2018} of one-dimensional systems with two interacting groups (majorities being advantaged in homophilic regimes and vice versa for minorities) can be understood using the second principle. 

\begin{figure}[h!]
\centering
\includegraphics[width=1.0\textwidth]{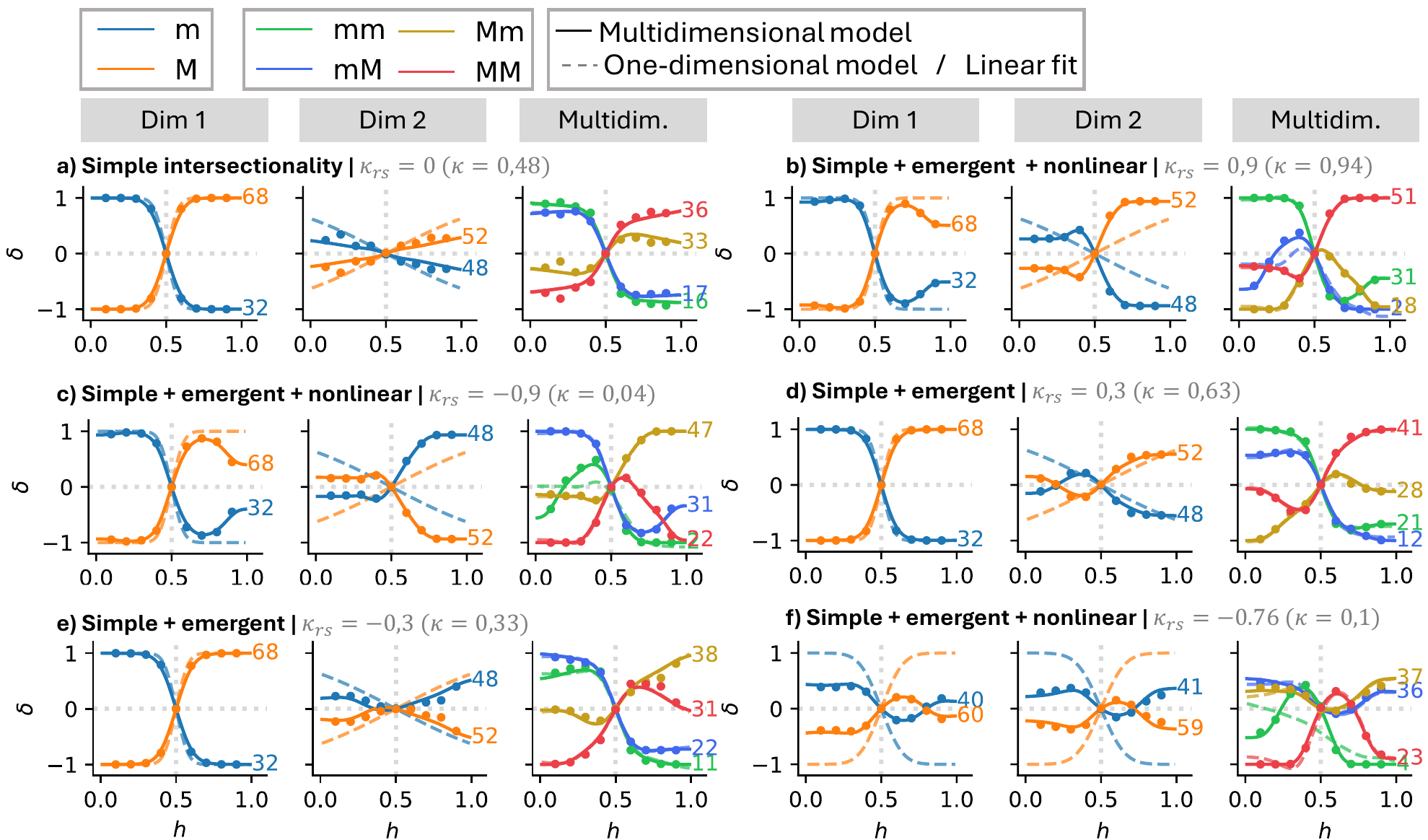}
\caption{\textbf{Inequality patterns in multidimensional systems}. Total disparity $\delta$ of one-dimensional groups (left and center plots of each panel) and multidimensional groups (rightmost plot) in a two-dimensional system with two categories per dimension, a ($m$)inority and a ($M$)ajority. The varying factor in each panel is the population distribution, while all other parameters are kept the same. The panels are labeled according to the types of intersectionality they best illustrate. We also show the values of the correlation parameter $\kappa\in[0,1]$ used to generate the population distribution and its rescaled version $\kappa_{rs}\in[-1,+1]$. The lines are labeled with the relative sizes of the groups as a percentage of the total population (they may sum up to more or less than 100 due to rounding errors).
Continuous lines are analytical values computed with Eqs. \eqref{eq:total_multi_ineq}  and \eqref{eq:total_onedim_ineq}. Circles show the results of 100 averaged simulations. In the one-dimensional plots, the dashed lines indicate the inequality that we would observe in a one-dimensional system with the same population distribution and one-dimensional preferences of that dimension. In the multidimensional plots, the dashed lines result from numerically fitting each multidimensional group's $\delta$ as a linear function of the two one-dimensional $\delta$. The gray dotted lines indicate the point of no inequality ($\delta = 0$) and neutral preference ($h=0.5$), respectively. }
\label{fig:examples_2D_inequality}
\end{figure}

In Fig.~\ref{fig:examples_2D_inequality}, we present the values of total disparity $\delta$ for one-dimensional and multidimensional groups as a function of parameter $\mathsf{h}$ of Eq. \eqref{eq:1dpref_tune_h}. Lines are analytical values computed with Eqs. \eqref{eq:total_multi_ineq}  and \eqref{eq:total_onedim_ineq}. Circles show the results of 100 averaged simulations. All computations have been performed for a network of 500 nodes. The only varying factor in each panel is the population distribution, which is controlled by the minority fractions ($f^1_{m}, f^2_{m}$), and correlation ($\kappa$). Correlation determines the relative group size of multidimensional groups and which one-dimensional groups are interdependent, which has a significant impact on whether they will enjoy advantages or suffer disadvantages. 
For convenience, we compute a rescaled version of $\kappa$ to obtain a correlation parameter $\kappa_{rs}$ bounded between $-1$ and $+1$, with $0$ indicating random mixing:

\begin{equation}
    \kappa_{rs}=
    \begin{cases} 
    \frac{\kappa-f^2_m}{1-f^2_m} &\text{if} \quad \kappa \geq f^2_m\\
    \frac{\kappa-f^2_m}{f^2_m} & \text{if}\quad  \kappa < f^2_m
    \end{cases} 
    \quad \text{;} \quad
    \kappa =
    \begin{cases}
    (1-f^2_m)\kappa_{rs} + f^2_m & \text{if} \quad \kappa_{rs} \geq 0\\
    f^2_m \kappa_{rs} + f^2_m & \text{if} \quad \kappa_{rs} < 0
    \end{cases}
\end{equation}
Each panel is labeled with the rescaled $\kappa_{rs}$ and original correlation parameter $\kappa$ used to generate the population distribution.

Within each panel in Fig.~\ref{fig:examples_2D_inequality}, the two leftmost plots correspond to the one-dimensional $\delta^d_{r^d}$ for each dimension and the rightmost plot, to the  $\delta_{\mathbf{r}}$ of multidimensional groups. For convenience, we have labeled each line with the relative size of the group as a percentage of the total population, so the label showing 32 next to the blue line of the top left plot indicates the smallest minority ($f^1_m=$32\% ) in panel \textbf{a}. In every plot, the solid lines and the circles show the inequality according to the multidimensional model. In the one-dimensional plots (left and center in each panel), the dashed lines indicate the inequality that we would observe in an analogous one-dimensional system with the same population distribution and one-dimensional preferences of that dimension (defined in Eq. \eqref{eq:1dpref_tune_h}). In multidimensional plots (most right columns), the dashed lines result from the numerical fitting of each multidimensional group $\delta_{\mathbf{r}}$ as a linear combination of the two one-dimensional $\delta_{r^1}, \delta_{r^2}
$.

Let us now explore the degree disparities found in this multidimensional system using the three notions of intersectionality we proposed above: simple intersectionality, emergent intersectionality, and nonlinear intersectionality.

\subsubsection{\emph{Simple intersectionality.}}
When attributes are not correlated, multidimensional and one-dimensional inequalities behave qualitatively similarly to one-dimensional systems, with larger groups being advantaged in the homophilic regime and smaller groups in the heterophilic regime. In Supp. Sec. \ref{sec:SI_understand_total_disp}, we rigorously prove that for any population distribution with uncorrelated attributes,  $\delta_{MM}\geq\delta_{Mm}\geq\delta_{mM}\geq\delta_{mm}$ if $\mathsf{h}>0.5$ and $\delta_{mm}\geq\delta_{mM}\geq\delta_{Mm}\geq\delta_{MM}$ if $\mathsf{h}<0.5$. However, as illustrated in Fig. \ref{fig:examples_2D_inequality}\textbf{a}, while the inequality patterns are qualitatively similar to those of analogous one-dimensional systems, they are in general not quantitatively the same (compare the continuous and dashed lines in the central plot). Since the attributes of this system are not correlated, we infer that the multidimensionality of interactions alone is enough to change the values of inequalities, leading to \emph{simple intersectionality}.

\subsubsection{\emph{Emergent intersectionality.}}
Attribute correlation induces unexpected patterns that are unobservable in one-dimensional systems. For example, in the multidimensional plot of panel \textbf{b}, we find that a smaller group ($Mm$ - 18\% size) has the advantage over a larger group ($mm$ - 31\%) in most of the homophilic range ($\mathsf{h}>0.5$). The reason is that, while both groups have similar sizes, the largest group ($MM$ - 51\%) has a relatively high preference for group $Mm$, while the groups with comparable preference for $mm$ are $mM$ and $Mm$, which are much smaller (2\% and 18\%). Another unexpected finding is that the two largest groups ($MM$ and $mm$) are the most privileged when the system is extremely heterophilic (we find an analogous pattern in panel \textbf{f}). The explanation in this case is that although they are the largest, they also have the highest connection preference towards each other when $\mathsf{h}<0.5$, placing them in an advantageous position.

By tuning the correlation value, we can generate situations where the regions that benefit the majority and the minority are inverted, as in the central plot of panel \textbf{c}. Or even more counterintuitive scenarios where a one-dimensional group is advantaged (or disadvantaged) at both ends of the homophily spectrum, as in the central plots of panels \textbf{d} and \textbf{e}. 


In panel \textbf{f} we show a particularly striking pattern where both one-dimensional minorities are simultaneously advantaged at both ends of the preference spectrum. However, at the same time, the multidimensional minority $mm$ is significantly disadvantaged in the same $\mathsf{h}$ range. Although seemingly paradoxical, this result is explained by the advantages of the mixed groups $mM$ and $Mm$, which are the largest (advantaged when $\mathsf{h}>0.5$) and connect to each other at a high rate in the heterophilic regime (when $\mathsf{h}<0.5$).

These complex and often counterintuitive patterns of privilege and marginalization are manifestations of \emph{emergent intersectionality}, as they are 
fundamentally distinct from those found in analogous one-dimensional binary systems. The intricate trends can be better understood by realizing that total disparities $\delta_{\mathbf{r}}$ are linear combinations of stochastic differences $\delta_{\mathbf{r},\mathbf{s}}$, as shown in Eq.~\eqref{eq:total_multi_ineq}. In Supp. Sec. \ref{sec:SI_understand_total_disp}, we discuss this matter in detail, demonstrating that \emph{emergent intersectionality} is only observable in the system when attributes are correlated.

Finally, although we can find emergent intersectionality for many parameter combinations, not all present the exotic patterns shown in Fig.~\ref{fig:examples_2D_inequality}. For instance, the one-dimensional groups of dimension 1 (the one with the smallest minority) present standard inequality patterns in almost every case. For a comprehensive characterization of multidimensional inequalities in this system, see Supp. Sec. \ref{sec:SI_inequality_maps}.

\subsubsection{\emph{Nonlinear intersectionality.}}
Another central aspect of intersectionality is the irreducibility of multidimensional inequalities to a simple addition of one-dimensional ones. 
To capture this notion of intersectionality, we say that multidimensional total disparities $\delta_\mathbf{r}$ are \emph{nonlinearly intersectional} if they \emph{can not} be expressed as a linear combination of the one-dimensional (dis)advantages $\delta^d_{r^d}$ experienced by their one-dimensional constituent groups $r^d$:

\begin{equation}
    \delta_{\mathbf{r}} \neq a_0(F) + \sum_d a_d(F) \delta^d_{r^d}
\end{equation}

Where the coefficients $a_1, a_2, \dots, a_D$  depend only on the population distribution $F$. The implication is that if total disparities are \emph{not intersectional}, once the $a_d(F)$ constants are found for a fixed population distribution $F$, $\delta_\mathbf{r}$ can always be computed from $\delta^d_{r^d}$ regardless of the connection preference values, and therefore multidimensional inequalities are a simple weighted combination of one-dimensional ones, contrary to the main notion of intersectionality. 

The inspection of the analytical expressions for degree inequalities that we have derived from the model (Eqs. \eqref{eq:pairwise_multi_ineq}-\eqref{eq:total_onedim_ineq}) suggests that, in general, we can not express multidimensional degree disparities (Eq.~\eqref{eq:total_multi_ineq}) as a function of one-dimensional ones (Eq.~\eqref{eq:total_onedim_ineq}). Therefore, multidimensional inequalities in this model are \emph{nonlinearly intersectional}.
To prove it, we just need to find counter-examples where multidimensional inequalities are not a linear combination of one-dimensional ones. In Methods, we examine one of those counter-examples analytically. Additionally, we have numerically computed linear regressions of $\delta_\mathbf{r}$ using the $\delta^d_{r^d}$ as the independent variables. The values predicted by the linear fits are shown as dashed lines in the rightmost plots of each panel. If inequalities were not intersectional, the dashed lines should \emph{perfectly align} with the continuous lines \emph{in every case}. Group $mM$ in panel \textbf{b} and group $mm$ in panels \textbf{c} and \textbf{f} are clear counter-examples where the $\delta_\mathbf{r}$ predicted by the linear fits significantly depart from the correct ones. From this computation, we conclude that the inequalities generated by the model are, in general, nonlinearly intersectional. 

In some panels, the dashed and continuous lines coincide almost perfectly. However, this should not be taken as evidence for the absence of nonlinear intersectionality,
 because finding a perfect linear fit is a necessary but not sufficient condition for declaring inequalities as non-intersectional. 
 Conversely, the examples where the linear regressions do not fit multidimensional inequalities are a sufficient but not necessary condition to identify inequalities as nonlinearly intersectional.

Interestingly, while multidimensional inequalities are not a linear combination of one-dimensional ones, the opposite is actually true: as shown in Methods, one-dimensional inequalities are always a linear combination of multidimensional ones. Taken together, these analyses underline the need to shift the perspective when studying inequalities in multidimensional systems. Multidimensional groups should be the fundamental unit of analysis, with the properties of one-dimensional groups considered as a combination of the characteristics of multidimensional ones. However, the one-dimensional group memberships should not be disregarded. While the experiences of a black woman are not a simple combination of those of black people and women, the experiences of black people are the combination of those of black men and black women (and many other multidimensional groups).



\section{Predicting real-world intersectional inequalities in networks}

In this section, 
we test whether the fitted model successfully predicts the group-level degree inequalities found in the data, which would be a strong indicator that the theory is correct and can be used to explore inequalities in hypothetical scenarios with alternative preferences and population distributions. To this end, we infer the model parameters from the data using the method developed in \cite{martin-gutierrezHiddenArchitectureConnections2024}.

We use the National Longitudinal Study of Adolescent Health (AddHealth) dataset \cite{moody_peer_2001}, which contains information about friendship relationships among high school students in the United States and about the students' sociodemographic attributes. We use a processed version of the dataset that includes directed friendship networks and the grade (between 7th and 12th), race (black, hispanic, asian, white, and mixed/other), and gender (boy or girl) of the students \cite{peixotoAdd_healthAdolescentHealth}.

\begin{figure}[h!]
\centering
\includegraphics[width=1.0\textwidth]{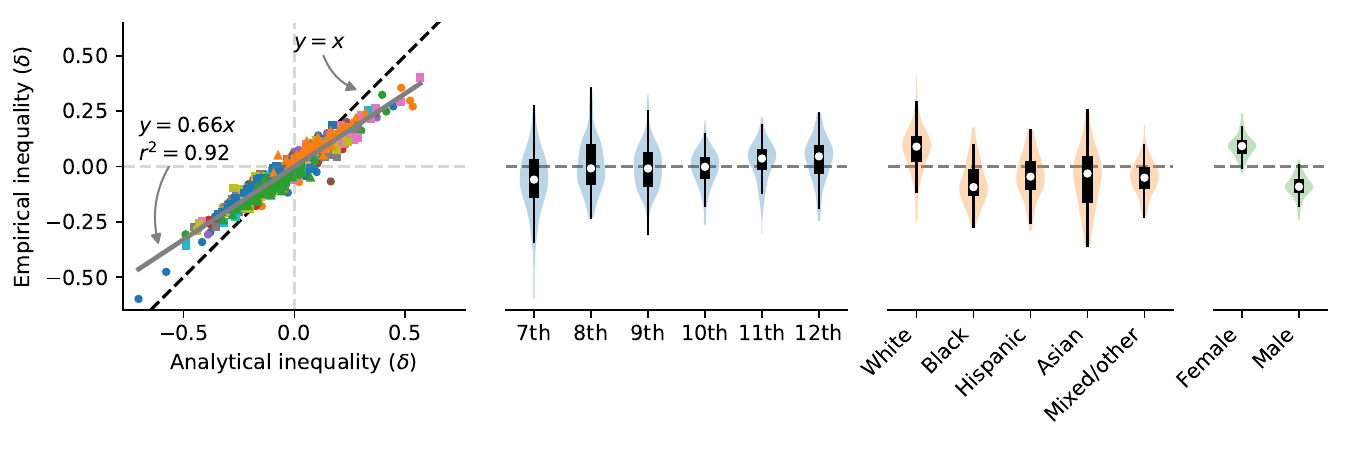}

\vspace{-0.2cm}

\includegraphics[width=1.0\textwidth]{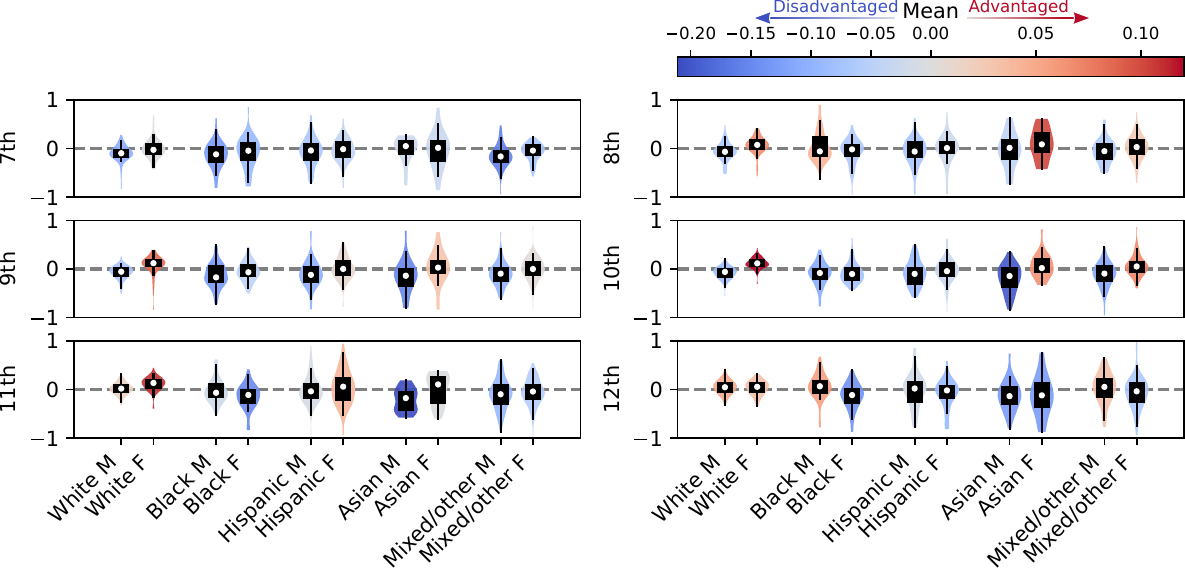}
\caption{\textbf{Inequalities of social capital in school friendship networks}. Inter-group in-degree inequalities as computed with Eq. \eqref{eq:inequality_metric} for the one-dimensional groups (top) and the multidimensional groups (bottom). The top left panel shows a comparison between empirical and predicted one-dimensional inequalities. The linear regression performed between the empirical and analytical inequalities is described by the equation $y = (0.659 \pm 0.007)x - (0.0006 \pm 0.0011)$. The y-axis in the bottom panels corresponds to the $\delta$ inequality metric, and the color to the mean value of the distribution of $\delta$.}
\label{fig:inequalities_data}
\end{figure}

We have computed the empirical in-degree disparities in the networks of the AddHealth dataset using Eq. \eqref{eq:inequality_metric}. Then, we have inferred the latent connection preferences in the dimensions of grade, race, and gender (see Supp. Fig.~\ref{fig:onedim_preferences}). Finally, we have computed the predicted inequalities by plugging the population fractions and inferred preferences into Eqs. \eqref{eq:pairwise_multi_ineq} and \eqref{eq:total_onedim_ineq}. In the top-left panel of Fig.~\ref{fig:inequalities_data}, we present the analytical versus the empirical inequalities for all one-dimensional groups in all networks. The marker shapes indicate different dimensions and the colors different attribute values within each dimension. They are nevertheless indiscernible because all points collapse into a tight region around the line $y = (0.659 \pm 0.007)x - (0.0006 \pm 0.0011)$. The correlation between theoretical and empirical results is remarkable, with the model explaining 92\% of the variance observed in the data according to Pearson's $r^2$. As a small caveat, the model slightly overestimates inequality. Since Eqs. \eqref{eq:pairwise_multi_ineq} and \eqref{eq:total_onedim_ineq} provide a near-perfect fit for the simulated data of Fig.~\ref{fig:examples_2D_inequality}, the source of the disparity between observed and theoretical inequalities may be unaccounted dynamics such as triadic closure, which has been observed to reduce network segregation (and therefore degree inequality) in some settings \cite{abebe_effect_2022}. While including this and other tie formation mechanisms, such as reciprocity, are natural extensions of the model that we will implement in future works, the extraordinary correspondence between empirical and theoretical inequalities observed in Fig.~\ref{fig:inequalities_data} demonstrates that inter-group inequalities are almost entirely explained by connection preferences. The incorporation of endogenous tie formation mechanisms will only introduce small linear corrections.

We can understand the empirical inequalities shown in the remaining top panels of Fig.~\ref{fig:inequalities_data} using what we learned from the analysis of synthetic networks in Fig.~\ref{fig:examples_2D_inequality}. For example, we observe a monotonously increasing trend in the $\delta$ measure for grades, indicating that a higher grade is always slightly advantaged over a lower one. The reason is that there is an aspirational linking trend where students from lower grades prefer to befriend those of higher grades more than vice versa (see Supp. Fig.~\ref{fig:onedim_preferences}). In the case of race, white people are the most privileged, probably due to them being homophilic and the largest group in many schools. Black students' disadvantage can be explained by the widespread race homophily and their smaller relative size (see Supp. Fig.~\ref{fig:group_sizes}). 
Finally, the gender degree disparity is due to an asymmetry in connection preferences, as boys have a higher preference for girls than girls for boys.

In the bottom panels of Fig.~\ref{fig:inequalities_data}, we show the in-degree disparities between the multidimensional groups. We can observe that while most groups are slightly disadvantaged on average, white girls tend to be advantaged in almost every grade. This is an expected result given the distribution of one-dimensional disparities. However, we also observe unexpected patterns: black boys have some advantage on average in 8th and 12th grades, while black students are the most disadvantaged in the race dimension and boys the most disadvantaged in the gender dimension. These nuances highlight the necessity of an intersectional perspective that addresses the inherent complexities of inequality. 

\section{Discussion}

{
The connection choices we make when building our social networks can create unanticipated social capital inequalities. 
This is especially true for multidimensional systems, where individuals at the intersection of several groups experience complex dynamics of privilege and marginalization.
To get a comprehensive understanding of intersectional inequalities in social networks, we have analyzed degree disparities in synthetic and real-world systems through the development of versatile network models and inference methods. 

We have operationalized three aspects of intersectional network inequality:
\textbf{Simple Intersectionality} captures the quantitative variations in inequality that arise when moving from a one-dimensional to a multidimensional system. In practice, this means that disparities may be amplified or mitigated when additional social dimensions are considered. For example, a social group that holds advantages in a one-dimensional context might experience intensified or reduced advantages in a multidimensional setting, though the overall pattern (advantage or disadvantage) may still be preserved.
\textbf{Emergent Intersectionality} refers to fundamentally new patterns of inequality that cannot be observed in one-dimensional systems. Intuitively, this aspect describes striking trends, such as a group being disadvantaged in a one-dimensional system but advantaged when more social dimensions operate in parallel. Another example is a multidimensional group experiencing significant disadvantage, even when each component group taken individually has an advantageous or neutral position.
\textbf{Nonlinear Intersectionality} denotes the irreducibility of multidimensional inequalities to linear combinations of one-dimensional ones.  In practical terms, this implies that multidimensional inequalities cannot be predicted by merely summing or weighting inequalities from individual dimensions.
Our work contributes to the growing body of literature around intersectionality that exists in other fields such as AI \cite{royMultidimensionalDiscriminationLaw2023}, law \cite{crenshawMappingMarginsIntersectionality1991}, or science of science \cite{doi:10.1073/pnas.2113067119}.
}
{
We have systematically studied a synthetic two-dimensional system with two categories per dimension (a minority group and a majority group), finding that attribute correlation (consolidation) plays a crucial role in shaping inequality patterns. Using the model's analytics, we rigorously demonstrate that without correlation, degree disparities mirror one-dimensional systems: majorities gain the advantage in homophilic regimes, while minorities benefit in heterophilic ones. However, even with uncorrelated attributes, we find \emph{simple intersectionality}, as inequalities take significantly different values than in analogous one-dimensional systems. 
}

{
Attribute correlation induces counterintuitive patterns of \emph{emergent intersectionality}. For example, for certain population distributions, the regions where majorities and minorities are advantaged are reversed. Other distributions result in one group holding the advantage at both ends of the homophily spectrum. We also find scenarios where individuals belonging to the minority group in either dimension are the most privileged, but those belonging to both minorities simultaneously are severely disadvantaged. To illustrate this last result, consider a hypothetical situation where women and Black people were each privileged in a certain context, but Black women —at the intersection of both minority groups— experienced significant disadvantage. Such complexity highlights the need for the holistic lens of intersectionality, as neither the one-dimensional nor the multidimensional perspectives in isolation are enough to fully understand the structures of privilege and marginalization. Finally, we have demonstrated that the multidimensional inequalities generated by our model are \emph{nonlinearly intersectional}; that is, they are not decomposable into linear combinations of one-dimensional inequalities.
}

{
While multidimensional \emph{preferences} in the network model are a simple multiplicative combination of one-dimensional preferences, multidimensional \emph{inequalities} are not. Conversely, we have shown that the inequalities of one-dimensional groups are linear combinations of the inequalities of multidimensional groups. This seemingly paradoxical result is quite natural: the experiences of a one-dimensional group are the combination of the experiences of the diverse multidimensional groups that compose it. From a practical perspective, it simply means that, while the experiences of Black women are not the addition of those of Black people and women (as emphasized by intersectionality theory), 
the Black experience is a mosaic comprised of the diverse perspectives of Black women, Black men, non-binary individuals, Black people across different socioeconomic backgrounds, and other diverse individuals at various categorical intersections.
}

{To validate the model's applicability to understand inequalities in real-world systems, we have compared the degree inequalities predicted by the theory to the empirical inequalities measured on school friendship networks, finding that they align remarkably well. 
Since the model correctly captures the general inter-group inequality trends, we conclude that inter-group inequalities are mainly driven by a combination of group size imbalances and biased connection preferences. 
}

{
To summarize, we have systematically operationalized intersectional inequalities in social networks, revealing how multidimensional interactions and attribute correlation impact social inequalities at multiple levels. Two key contributions of our work are the derivation of closed-form analytical expressions for the inequalities predicted by the network model and their validation with empirical data. These methodological advances will pave the way to better understand intersectional inequalities in online social networks and algorithms.
}

\section{Methods}

\subsection{Analytical computation of in-degree inequalities}
To derive an analytical expression for the inter-group degree inequalities $\delta_{\mathbf{r},\mathbf{s}}$ generated by the network model \cite{martin-gutierrezHiddenArchitectureConnections2024}, we use a Poissonian approximation for the in-degree distribution of a multidimensional group $\mathbf{r}$ \cite{karrerStochasticBlockmodelsCommunity2011a}:

\begin{equation}
    P(k_{i \in \mathbf{r}}) = \frac{(\lambda_\mathbf{r})^k e^{-\lambda_\mathbf{r}}}{k!}
\end{equation}

We now need to find the $\lambda_\mathbf{r}$ parameter, which corresponds to the expected in-degree. There are $F_{\mathbf{s}} N$  nodes of type $\mathbf{s}$, and they connect to nodes of type $\mathbf{r}$ at a rate given by their multidimensional preference $H_{\mathbf{s}, \mathbf{r}}$, so the total expected number of incoming links to an average node of group $\mathbf{r}$ is

\begin{equation}
    \lambda_\mathbf{r} = N \sum_{\mathbf{s}} F_\mathbf{s} H_{\mathbf{s},\mathbf{r}}
\end{equation}
We can now compute $P(k_{i \in \mathbf{r}} > k_{j \in \mathbf{s}})$ and $P(k_{i \in \mathbf{r}} < k_{j \in \mathbf{s}})$
using the distribution of the difference of two Poissonian variables, called Skellam distribution \cite{skellamFrequencyDistributionDifference1946}:

\begin{equation}
    P(k_{i \in \mathbf{r}} - k_{j \in \mathbf{s}} = \Delta) = 
    e^{-(\lambda_\mathbf{r}+\lambda_\mathbf{s})} 
    \left( \frac{\lambda_{\mathbf{r}}}{\lambda_{\mathbf{s}}} \right)^{\frac{\Delta}{2}} I_{\Delta} (2\sqrt{ \lambda_{\mathbf{r}} \lambda_{\mathbf{s}} } )
\end{equation}

Where $I_\Delta (\cdot)$ is the modified Bessel function of the first kind of order $\Delta$. The probability that we need to compute is $P(k_{i \in \mathbf{r}} > k_{j \in \mathbf{s}})=P(k_{i \in \mathbf{r}} - k_{j \in \mathbf{s}}>0) = P(k_{i \in \mathbf{r}} - k_{j \in \mathbf{s}} \geq 1)$, where in the last expression we simply take into account that degrees are discrete quantitites. Considering the expression for Skellam's PMF:

\begin{equation}
\label{eq:methods_skellam_CDF}
    P(k_{i \in \mathbf{r}} > k_{j \in \mathbf{s}}) = P(k_{i \in \mathbf{r}} - k_{j \in \mathbf{s}} \geq 1) = P(\Delta \geq 1)=  e^{-(\lambda_\mathbf{r}+\lambda_\mathbf{s})}
    \sum_{\Delta=1}^{\infty}
    \left( \frac{\lambda_{\mathbf{r}}}{\lambda_{\mathbf{s}}} \right)^{\frac{\Delta}{2}} I_{\Delta} (2\sqrt{ \lambda_{\mathbf{r}} \lambda_{\mathbf{s}} } )
\end{equation}

This expression can be written more compactly using generalized Marcum Q-functions $Q_\nu(a,b)$ thanks to the following property \cite{annamalaiSimpleExponentialIntegral2008}:

\begin{equation}
\label{eq:methods_marcumq}
    1-Q_\nu(a,b) = e^\frac{(a^2+b^2)}{2} \sum_{\alpha=\nu}^\infty \left( \frac{b}{a} \right)^\alpha I_\alpha (a,b)   
\end{equation}

By comparing Eqs. \eqref{eq:methods_skellam_CDF} and \eqref{eq:methods_marcumq}, we realize that if we set $a=\sqrt{2\lambda_{\mathbf{s}}},b=\sqrt{2\lambda_{\mathbf{r}}},\nu =1$, we have that:

\begin{equation}
\label{eq:methods_upper}
    P(k_{i \in \mathbf{r}} > k_{j \in \mathbf{s}})=P(k_{i \in \mathbf{r}} - k_{j \in \mathbf{s}} \geq 1) = P(\Delta \geq 1)=  1-Q_1(\sqrt{2\lambda_{\mathbf{s}}},\sqrt{2\lambda_{\mathbf{r}}}) 
\end{equation}
By symmetry, 

\begin{equation}
\label{eq:methods_lower}
    P(k_{i \in \mathbf{r}} < k_{j \in \mathbf{s}}) =  1-Q_1(\sqrt{2\lambda_{\mathbf{r}}},\sqrt{2\lambda_{\mathbf{s}}})
\end{equation}

Plugging Eqs. \eqref{eq:methods_upper} and \eqref{eq:methods_lower} into the expression for $\delta_{\mathbf{r},\mathbf{s}}$ of Eq. \eqref{eq:inequality_metric_pairwise}, we obtain:

\begin{equation}
\label{eq:methods_multidim_pairwise_ineq}
    \delta_{\mathbf{r}, \mathbf{s}} =  P(k_{i \in \mathbf{r}} > k_{j \in \mathbf{s}}) - P(k_{i \in \mathbf{r}} < k_{j \in \mathbf{s}})= Q_1(\sqrt{2 \lambda_{\mathbf{r}} }, \sqrt {2 \lambda_{\mathbf{s}}}) - Q_1(\sqrt{2 \lambda_{\mathbf{s}} }, \sqrt {2 \lambda_{\mathbf{r}}})
\end{equation}

In practice, we use SciPy's implementation of Skellam's distribution to compute the probabilities.

A useful property of the stochastic difference $\delta$ is that, if $R$ and $S$ are composed of disjoint collections of subsets ($R=r_1 \cup r_2 \cup \dots | r_{\alpha_1} \cap r_{\alpha_2} = \emptyset \; \forall \; \alpha_1 \neq \alpha_2$, and the same for $S$),  the metric $\delta_{R,S}$ can be written as a linear combination of the pairwise inequalities $\delta_{r_\alpha, s_\beta}$ between the subsets $r_\alpha$ and $s_\beta$. The reason is that the process for computing $\delta $ can be imagined as first picking a random node from group $R$, which will belong to a subset $r_\alpha$ with a probability proportional to its size $|r_\alpha| / |R|$. Secondly, we pick a random node from $S$ that will belong to a subset $s_\beta$ with probability $|s_\beta| / |S|$. The probability that the first node's in-degree is larger than the second is $P(k_{i\in r_\alpha} > k_{j \in r_\beta})$. Therefore, the joint probability of picking a random node from $r_\alpha$, another from $s_\beta$, and that the first has higher degree is $\frac{|r_\alpha|}{|R|} \frac{|s_\beta|}{|S|} P(k_{i\in r_\alpha} > k_{j \in r_\beta})$. By summing over all pairs of subsets $(r_\alpha, s_\beta)$ from $R$ and $S$, we obtain 

\begin{equation}
    P(k_{i \in R} > k_{j \in S}) = \sum_{\alpha, \beta} \frac{|r_\alpha|}{|R|} \frac{|s_\beta|}{|S|} P(k_{i\in r_\alpha} > k_{j \in r_\beta})
\end{equation}
Doing the same for  $P(k_{i \in R} < k_{j \in S})$ we arrive at the following expression for $\delta_{R,S}$:

\begin{equation}
    \delta_{R,S} = \sum_{\alpha, \beta} \frac{|r_\alpha|}{|R|} \frac{|s_\beta|}{|S|} \delta_{r_\alpha, s_\beta} =\frac{1}{|R| |S|} \sum_{\alpha, \beta} |r_\alpha| |s_\beta| \delta_{r_\alpha, s_\beta}
    \label{eq:disparity_lin_comb}
\end{equation}


Using this property, we can compute the total disparity of a multidimensional group $\mathbf{r}$ as:

\begin{equation}
    \delta_\mathbf{r} = \frac{1}{1- F_\mathbf{r}} \sum_{\boldsymbol{\sigma} \neq \mathbf{r}} \delta_{\mathbf{r}, \boldsymbol{\sigma}} F_{\boldsymbol{\sigma}} %
    \label{eq:methods_total_disparity_multidim}
\end{equation}
One-dimensional groups are a union of several multidimensional ones ($r^d = \bigcup\limits_{\rho^d = r^d}  \boldsymbol{\rho}$). For example, in the system illustrated in the top panels of Fig.~\ref{fig:model}, the one-dimensional group $\female$ is a union of the multidimensional groups $(\female,A)$, $(\female,B)$, and $(\female,C)$, while group $B$ is the union of $(\female,B)$ and $(\mars,B)$. Therefore, we can also use Eq. \eqref{eq:disparity_lin_comb} to compute the pairwise disparity $\delta^d_{r^d, s^d}$ between two one-dimensional groups $(r^d, s^d)$ or the total disparity $\delta^d_{r^d}$ of a given one-dimensional group $r^d$ with respect to the rest as follows:

\begin{align}
\label{eq:methods_onedim_pairwise_ineq}
    \delta^d_{r^d, s^d} = &%
    \frac{1}{f^d_{r^d} f^d_{s^d}}%
    \sum_{\rho^d=r^d, \sigma^d=s^d} F_{\boldsymbol{\rho}} F_{\boldsymbol{\sigma}} \delta_{\boldsymbol{\rho}, \boldsymbol{\sigma}} \\
\label{eq:methods_onedim_total_disparity}
    \delta^d_{r^d} = &
    \frac{1}{f^d_{r^d} (1-f^d_{r^d})}%
    \sum_{\rho^d=r^d, \sigma^d \neq r^d} F_{\boldsymbol{\rho}} F_{\boldsymbol{\sigma}} \delta_{\boldsymbol{\rho}, \boldsymbol{\sigma}}
\end{align}

By plugging the multidimensional stochastic difference of Eq. \eqref{eq:methods_multidim_pairwise_ineq} into Eqs. \eqref{eq:methods_total_disparity_multidim}-\eqref{eq:methods_onedim_total_disparity}, we can obtain the pairwise one-dimensional stochastic difference and the one- and multidimensional total disparity values directly from the model parameters $F$ and $h^d$:

\begin{equation}
\label{eq:pairwise_onedim_ineq_full}
\begin{split}
\delta^d_{r^d, s^d} = %
    \frac{1}{f^d_{r^d} f^d_{s^d}}%
    \sum_{\rho^d=r^d, \sigma^d=s^d} 
    F_{\boldsymbol{\rho}} F_{\boldsymbol{\sigma}} 
    &
    \left[ 
    Q_1 \left(
    \sqrt{
    2 N \sum_{\boldsymbol{\tau}} 
    F_{\boldsymbol{\tau}} 
    \prod_{d=1}^D h^d_{\tau^d,\rho^d} 
    }, 
    \sqrt {
    2 N \sum_{\boldsymbol{\tau}} 
    F_{\boldsymbol{\tau}} 
    \prod_{d=1}^D h^d_{\tau^d,\sigma^d} 
    } \right)
     \right.
    \\
    & \left.
    - Q_1 \left(
    \sqrt{
    2 N \sum_{\boldsymbol{\tau}} 
    F_{\boldsymbol{\tau}}
    \prod_{d=1}^D h^d_{\tau^d,\sigma^d} 
    }, 
    \sqrt {
    2 N \sum_{\boldsymbol{\tau}} 
    F_{\boldsymbol{\tau}} 
    \prod_{d=1}^D h^d_{\tau^d,\rho^d} 
    } \right) \right]
\end{split}
\end{equation}

\begin{equation}
\begin{split}
\delta_\mathbf{r} = 
\frac{1}{1- F_\mathbf{r}} 
\sum_{\boldsymbol{\sigma} \neq \mathbf{r}} 
F_{\boldsymbol{\sigma}} 
&\left[ 
    Q_1 \left(
    \sqrt{
    2 N \sum_{\boldsymbol{\tau}} 
    F_{\boldsymbol{\tau}} 
    \prod_{d=1}^D h^d_{\tau^d,r^d} 
    }, 
    \sqrt {
    2 N \sum_{\boldsymbol{\tau}} 
    F_{\boldsymbol{\tau}} 
    \prod_{d=1}^D h^d_{\tau^d,\sigma^d} 
    } \right) 
     \right.
    \\
    & \left.
    - Q_1 \left(
    \sqrt{
    2 N \sum_{\boldsymbol{\tau}} 
    F_{\boldsymbol{\tau}}
    \prod_{d=1}^D h^d_{\tau^d,\sigma^d} 
    }, 
    \sqrt {
    2 N \sum_{\boldsymbol{\tau}} 
    F_{\boldsymbol{\tau}} 
    \prod_{d=1}^D h^d_{\tau^d,r^d} 
    } \right)
    \right]
\end{split}
\end{equation}

\begin{equation}
\label{eq:total_onedime_ineq_full}
\begin{split}
    \delta^d_{r^d} =
    \frac{1}{f^d_{r^d } (1-f^d_{r^d})}%
    \sum_{\rho^d=r^d, \sigma^d \neq r^d} 
    F_{\boldsymbol{\rho}} F_{\boldsymbol{\sigma}} 
    &\left[
    Q_1 \left(
    \sqrt{
    2 N \sum_{\boldsymbol{\tau}} 
    F_{\boldsymbol{\tau}} 
    \prod_{d=1}^D h^d_{\tau^d,\rho^d} 
    }, 
    \sqrt {
    2 N \sum_{\boldsymbol{\tau}} 
    F_{\boldsymbol{\tau}} 
    \prod_{d=1}^D h^d_{\tau^d,\sigma^d} 
    } \right) 
    \right.
    \\
    & \left.
    - Q_1 \left(
    \sqrt{
    2 N \sum_{\boldsymbol{\tau}} 
    F_{\boldsymbol{\tau}}
    \prod_{d=1}^D h^d_{\tau^d,\sigma^d} 
    }, 
    \sqrt {
    2 N \sum_{\boldsymbol{\tau}} 
    F_{\boldsymbol{\tau}} 
    \prod_{d=1}^D h^d_{\tau^d,\rho^d} 
    } \right) 
    \right]
\end{split}
\end{equation}

\subsection{Counter-example demonstrating nonlinear intersectionality 
}

Here we show that the total disparity $\delta_{\mathbf{r}}$ of a multidimensional group $\mathbf{r}$ can not in general be expressed as a linear combination of its constituent one-dimensional groups. For this example, we consider the same 2D system analyzed throughout the paper, with two categories per dimension ($m$inority and $M$ajority). Let us focus on the total disparity of group $m,m$.
By expanding the corresponding expression from Eq.~\eqref{eq:methods_total_disparity_multidim} we obtain:

\begin{equation}
    \delta_{(m,m)} = \frac{1}{1-F_{(m,m)}} \left[ F_{(M,M)} \delta_{(m,m),(M,M)} + F_{(M,m)} \delta_{(m,m),(M,m)} + F_{(m,M)} \delta_{(m,m),(m,M)} \right]
    \label{eq:methods_mm_disparity}
\end{equation}

The total degree disparities of the two one-dimensional groups are:

\begin{align}
\label{eq:methods_m1_disparity}
    \begin{aligned}
    \delta^1_m = 
    \frac{1}{f^1_m(1-f^1_m)}
    &\left[ 
    \mathcolorbox{green}{ F_{(m,m)} F_{(M,m)} \delta_{(m,m),(M,m)}} +\right.& 
    \mathcolorbox{green}{F_{(m,m)}F_{(M,M)}\delta_{(m,m),(M,M)} } +\\
    & \mathcolorbox{yellow}{F_{(m,M)}F_{(M,m)}\delta_{(m,M),(M,m)}} +&
    \left. \mathcolorbox{red}{F_{(m,M)}F_{(M,M)}\delta_{(m,M),(M,M)} }
    \right]
    \end{aligned}
    \\
    \label{eq:methods_m2_disparity}
    \begin{aligned}
    \delta^2_m = 
    \frac{1}{f^2_m(1-f^2_m)} 
    &\left[
    \mathcolorbox{green}{F_{(m,m)}F_{(m,M)}\delta_{(m,m),(m,M)} } +\right.&
    \mathcolorbox{green}{F_{(m,m)}F_{(M,M)}\delta_{(m,m),(M,M)} } +\\
    &\mathcolorbox{yellow}{F_{(M,m)}F_{(m,M)}\delta_{(M,m),(m,M)}} +&
    \left.\mathcolorbox{red}{F_{(M,m)}F_{(M,M)}\delta_{(M,m),(M,M)} }
    \right] \\
    \end{aligned}
\end{align}

For $\delta_{(m,m)}$ to be a linear combination of the $\delta^d_{r^d}$ in all cases; that is:

\begin{equation}
    \delta_{(m,m)}=a_0+a_1 \delta^1_m+a_2 \delta^2_m,
    \label{eq:methods_mm_lin_comb}
\end{equation}

the $a$ coefficients should be independent of the specific values of the deltas $\delta_{\mathbf{r},\mathbf{s}}$. To achieve this in the general case, the following necessary (but not sufficient) conditions must be fulfilled:

\begin{enumerate}
    \item The $\delta_{\mathbf{r},\mathbf{s}}$ in Eqs. \eqref{eq:methods_m1_disparity}-\eqref{eq:methods_m2_disparity} must be the same as those appearing in Eq. \eqref{eq:methods_mm_disparity}.
    \item If some $\delta_{\mathbf{r},\mathbf{s}}$ appear only in Eqs. \eqref{eq:methods_m1_disparity}-\eqref{eq:methods_m2_disparity}  and not in Eq. \eqref{eq:methods_mm_disparity}, they must be present in both Eqs. \eqref{eq:methods_m1_disparity}-\eqref{eq:methods_m2_disparity}  and cancel each other when Eqs. \eqref{eq:methods_m1_disparity}-\eqref{eq:methods_m2_disparity} are plugged into Eq. \eqref{eq:methods_mm_lin_comb}.
\end{enumerate}
 
The terms in Eqs. \eqref{eq:methods_m1_disparity}-\eqref{eq:methods_m2_disparity} that fulfill condition 1 are highlighted in green and the terms that \emph{might} fulfill condition 2, in yellow (recall that $\delta_{\mathbf{r},\mathbf{s}} = - \delta_{\mathbf{s},\mathbf{r}}$). However, some terms fulfill neither (highlighted in red), which makes it impossible to find the $a$ coefficients for Eq.~\eqref{eq:methods_mm_lin_comb} in the general case (for any value of the $\delta_{\mathbf{r},\mathbf{s}}$ ).

\subsection{Inequalities experienced by one-dimensional groups are a linear combination of those experienced by multidimensional groups}

We have shown with numerical experiments and an analytical counter-example that the total disparity of multidimensional groups is not a linear combination of the total disparity experienced by their component one-dimensional groups. Here, we show that the converse is true. Let us recall that the one-dimensional total disparity is:

\begin{equation}
\label{eq:methods_onedim_total_disparity2}
        \delta^d_{r^d} = 
    \frac{1}{f^d_{r^d} (1-f^d_{r^d})}%
    \sum_{\rho^d=r^d, \sigma^d \neq r^d} F_{\boldsymbol{\rho}} F_{\boldsymbol{\sigma}} \delta_{\boldsymbol{\rho}, \boldsymbol{\sigma}}
\end{equation}

And let us label as $\boldsymbol{\rho}_1, \boldsymbol{\rho}_2, \dots, \boldsymbol{\rho}_Z$ all the vectors whose $d$ component is $r^d $ ($\rho^d=r^d$). Notice that there are in total $Z= \prod_{\tau\neq d}v_\tau$  of those vectors, where $v_\tau$ is the number of attribute values in dimension $\tau$. We can rewrite Eq.~\eqref{eq:methods_onedim_total_disparity2} as:

\begin{equation}
        \delta^d_{r^d} =
    \frac{1}{f^d_{r^d}(1-f^d_{r^d})} 
    \left[
    F_{\boldsymbol{\rho}_1} \sum_{\boldsymbol{\sigma} \neq \boldsymbol{\rho}_1} F_{\boldsymbol{\sigma}} \delta_{\boldsymbol{\rho}_1, \boldsymbol{\sigma}}
    +
     F_{\boldsymbol{\rho}_2} \sum_{\boldsymbol{\sigma} \neq \boldsymbol{\rho}_2} F_{\boldsymbol{\sigma}} \delta_{\boldsymbol{\rho}_2, \boldsymbol{\sigma}}
     +
     \dots
     +
      F_{\boldsymbol{\rho}_Z} \sum_{\boldsymbol{\sigma} \neq \boldsymbol{\rho}_Z} F_{\boldsymbol{\sigma}} \delta_{\boldsymbol{\rho}_Z, \boldsymbol{\sigma}}
      \right]
\end{equation}

Now, recalling Eq.~\eqref{eq:methods_total_disparity_multidim}, we can write this as a function of the total disparity $\delta_{\boldsymbol{\rho}_z}$ of the multidimensional groups $\boldsymbol{\rho}_z$ with $\rho^d_z = r^d$:

\begin{equation}
    \delta^d_{r^d} = \frac{1}{f^d_{r^d}(1-f^d_{r^d})} \sum_{z=1}^{z=Z} F_{\boldsymbol{\rho}_z} (1-F_{\boldsymbol{\rho}_z}) \delta_{\boldsymbol{\rho}_z}
\end{equation}
Proving that the total disparity $\delta^d_{r^d}$ of one-dimensional group $r^d = \bigcup_{z=1}^Z \boldsymbol{\rho}_z$ is a linear combination of the total disparities $\delta_{\boldsymbol{\rho}_z}$ of the multidimensional groups sharing the same $r^d$ value in dimension $d$, with linear coefficients $a_z = \frac{1}{f^d_{r^d}(1-f^d_{r^d})}  F_{\boldsymbol{\rho}_z} (1-F_{\boldsymbol{\rho}_z})$.

\section*{Acknowledgments}

This research work was funded by the European Union under the Horizon Europe MAMMOth project, Grant Agreement ID: 101070285, and by the Austrian Research Promotion Agency (FFG) under project No. 873927.

\section*{Data and code statement}
The data analyzed in this paper is available in the following repository: \url{https://networks.skewed.de/net/add_health}.
The code for reproducing the paper results is available at \url{https://github.com/CSHVienna/intersectional_network_inequalities_paper}.
We have also developed a python package called \texttt{multisoc} to infer latent preferences from data, simulate and analyze multidimensional networks, and characterize intersectional degree inequalities, available at \url{https://github.com/CSHVienna/multisoc}.

\bibliography{bibliography/zotero}
\bibliographystyle{splncs04}

\newpage
\begin{center}
\textbf{\LARGE Supplementary Information}
\end{center}
\setcounter{equation}{0}
\setcounter{figure}{0}
\setcounter{table}{0}
\setcounter{page}{1}
\setcounter{section}{0}
\makeatletter
\renewcommand{\theequation}{S\arabic{equation}}
\renewcommand{\thesection}{S\arabic{section}}
\renewcommand{\thefigure}{S\arabic{figure}}
\renewcommand{\thetable}{S\arabic{table}}

\section{Data description}
\label{sec:appendix_data}

The AddHealth dataset was built through a social survey carried out in 84 communities in 1994-95. Some of these communities had one high school and others two, usually split into junior high and high school. Each student was given a paper-and-pencil questionnaire and a copy of a roster listing every student in the school (or schools, if there were two). The name generator asked about five male and five female friends separately. The question was, "List your closest (male/female) friends. List your best (male/female) friend first, then your next best friend, and so on. (girls/boys) may include (boys/girls) who are friends and (boy/girl) friends." For each friend named, the student was asked to check off whether he/she participated in any of five activities with the friend: going to her house, meeting after school to hang out, spending time together during the weekend, talking about a problem in the last seven days, and talking on the phone in the last seven days. For our analyses, we have used all the friendship nominations regardless of the number or type of activities shared by the students.

One potential issue given that the question asks about male and female friends separately is that gender preference may be underestimated. However, students rarely provide a full list of 10 friends (out of all respondents, only 3\% do). Furthermore, we have found gender homophily and a clear asymmetry in gender preferences, so despite the question's wording, we are able to capture gender-based connection biases.

To prepare the data for the analysis, we removed all the nodes with unreported attributes. We also removed one-dimensional groups with fewer than 20 individuals to reduce the noise of the results. Our results are very robust to this filtering, as the inferred preferences for the groups that were not removed remain almost unchanged. Finally, we only analyzed schools with more than 100 students remaining in the network after filtering and with more than one gender and race category.  As a result, 41880 students from 70 communities were included in our study. We have analyzed each of the 70 communities independently.

\newpage
\section{Distribution of group sizes}

\begin{figure}[h!]
\centering
\includegraphics[width=\textwidth]{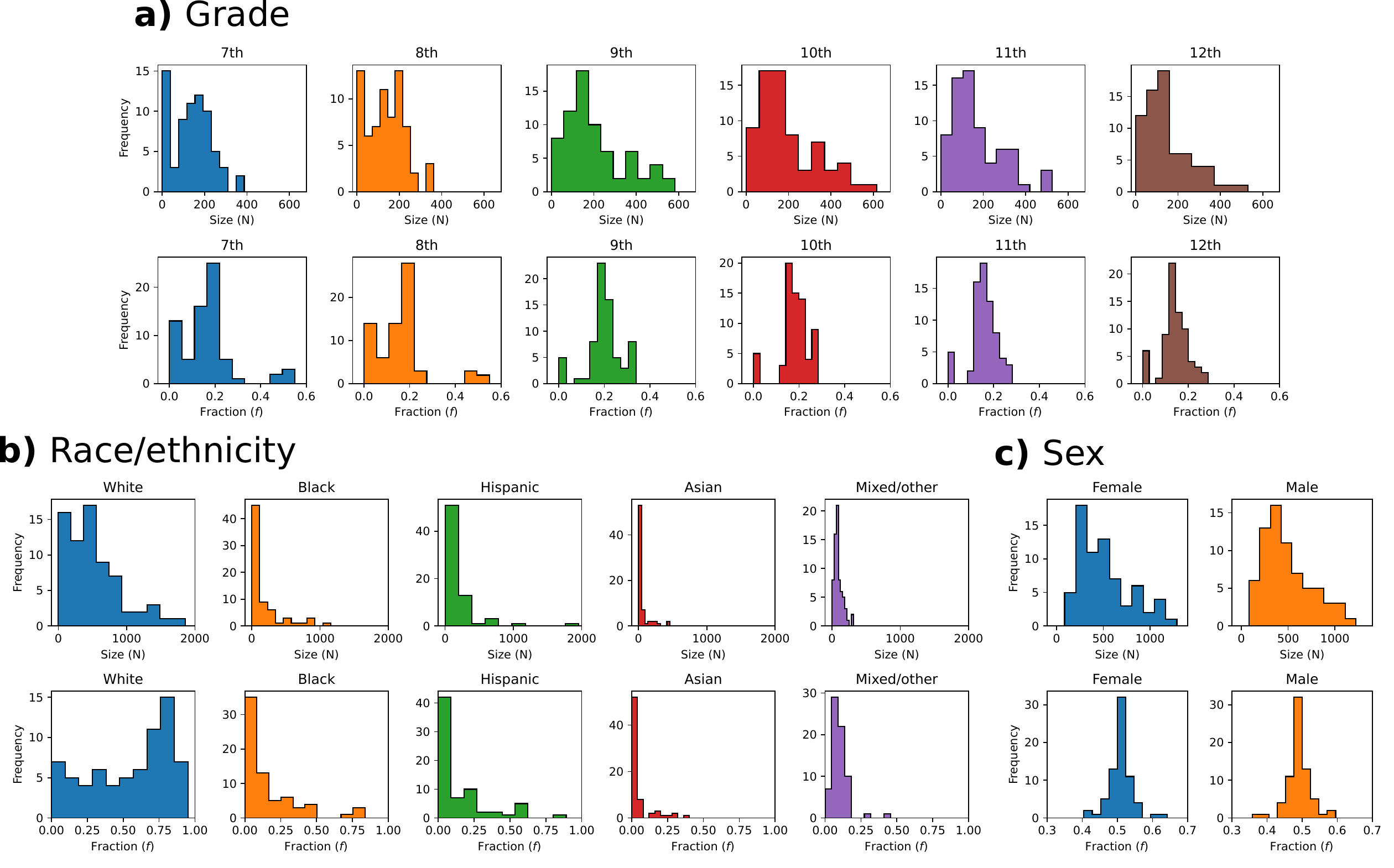}
\caption{\textbf{Distribution of group sizes in the AddHealth networks}. In each panel the top plots show the absolute size distribution in numbers of nodes and the bottom plots the relative size distributions as a fraction of the total population in each network.}
\label{fig:group_sizes}
\end{figure}

\section{Connection preferences in high school friendship networks}
\label{sec:app_connection_preferences}

We have inferred the latent one-dimensional connection preferences between high school students of the AddHealth dataset using a recently developed network model \cite{martin-gutierrezHiddenArchitectureConnections2024}. To make 1D preferences comparable across different networks and easier to interpret, we have normalized them by dividing each row of the one-dimensional preference matrices by the diagonal term; that is, the in-group preference of homophily $\frac{h_{r,s}}{h_{r,r}}$.

\begin{figure}[h!]
\centering
\includegraphics[width=0.9\textwidth]{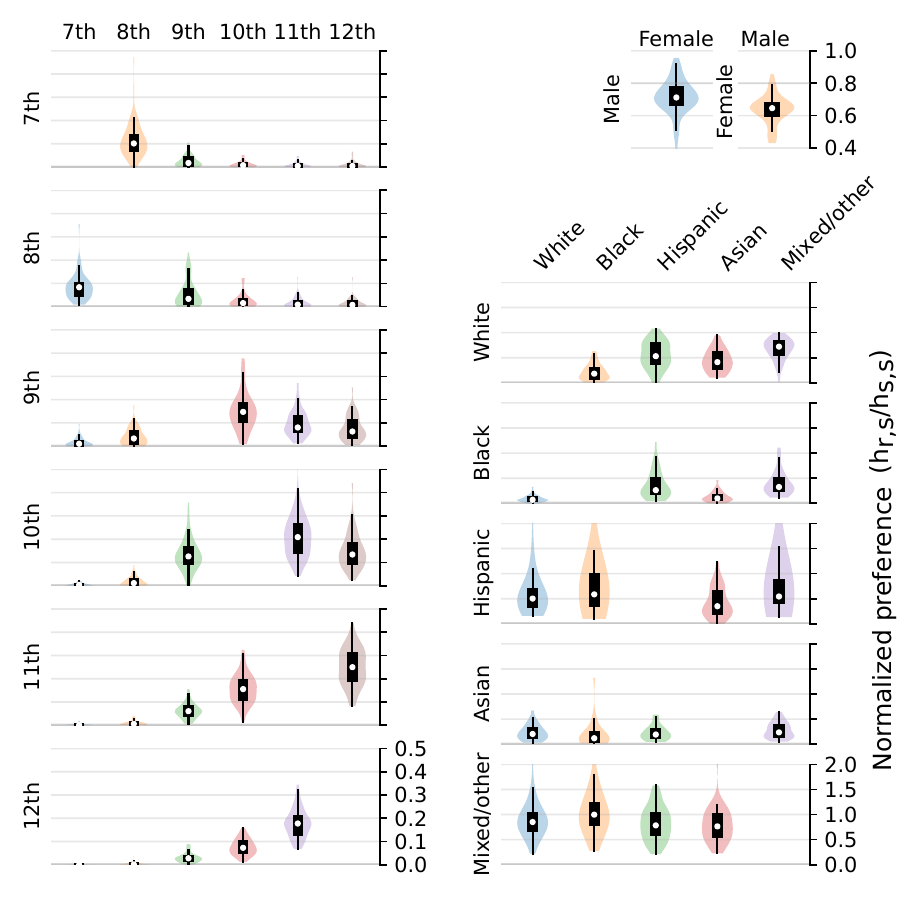}
\caption{\textbf{One-dimensional connection preferences in high school friendships (AddHealth)}. Preferences are obtained for each one-dimensional group in each school and then normalized by dividing by the in-group preference. Therefore, the normalized in-group preference is 1 by construction; we do not show it in this figure. Values lower than 1 in all the out-group preferences indicate a consistent homophily, while a value higher than 1 indicates heterophily for that particular group. Each violin plot and box plot combination represents the distribution of values obtained from fitting all the considered schools in the AddHealth dataset.}
\label{fig:onedim_preferences}
\end{figure}

In Fig.~\ref{fig:onedim_preferences} we show the distribution of preferences for the dimensions of grade, race, and gender. Each violin plot represents the distribution of the preference values obtained for all the considered schools. Since all preferences are normalized by in-group preference, a value higher than 1 indicates a higher tendency to connect to that group than to the in-group and vice versa. In the grade dimension, all preferences are lower than 1, indicating strong homophilic tendencies, as expected. In this dimension, preference is partially confounded with opportunity, as students naturally have more opportunities to interact with other schoolmates from their own grade. Nevertheless, the relative preferences show a remarkably regular and informative trend. Students prefer to connect with schoolmates of neighboring grades; however, this preference is not symmetric but aspirational. They prefer schoolmates of higher grades over those of lower grades. In the race dimension preferences show higher variation, sometimes reaching values above the in-group baseline, with students from different races presenting significantly different behaviors. Asian students are the most homophilic overall, while Mixed and Hispanic students are the most receptive to cross-group connections. The latter are likely forming bridges across homogeneous groups, as some cross-group preferences are particularly low. For example, we can observe a mutual avoidance between Black and White students, which can be understood in the wider context of racial interactions in the United States. Finally, while all students are homophilic in terms of gender, boys show a slightly higher relative preference towards girls than the other way around.

\section{Understanding the trends of multidimensional degree disparities}
\label{sec:SI_understand_total_disp}

As stated in Eq.~\eqref{eq:methods_total_disparity_multidim}, the total disparity experienced by a multidimensional group $\delta_{\mathbf{r}}$ is a linear combination of the stochastic differences between the focal group $\mathbf{r}$ and the rest:

\begin{equation}
    \delta_\mathbf{r} = \frac{1}{1- F_\mathbf{r}} \sum_{\boldsymbol{\sigma} \neq \mathbf{r}} \delta_{\mathbf{r}, \boldsymbol{\sigma}} F_{\boldsymbol{\sigma}}
    \label{eq:SI_ineq_linear_comb}
\end{equation}

Therefore, to understand the values taken by this metric e.g. in the compuations shown in Fig.~\ref{fig:examples_2D_inequality}, we need to examine the $\delta_{\mathbf{r},\mathbf{s}}$, which in turn depend on the expected in-degree of the multidimensional groups:

\begin{equation}
    \delta_{\mathbf{r}, \mathbf{s}} =  P(k_{i \in \mathbf{r}} > k_{j \in \mathbf{s}}) - P(k_{i \in \mathbf{r}} < k_{j \in \mathbf{s}})= Q_1(\sqrt{2 \lambda_{\mathbf{r}} }, \sqrt {2 \lambda_{\mathbf{s}}}) - Q_1(\sqrt{2 \lambda_{\mathbf{s}} }, \sqrt {2 \lambda_{\mathbf{r}}})
    \label{eq:SI_multidim_1v1_ineq}
\end{equation}

\begin{figure}[h!]
\centering
\includegraphics[width=1.0\textwidth]{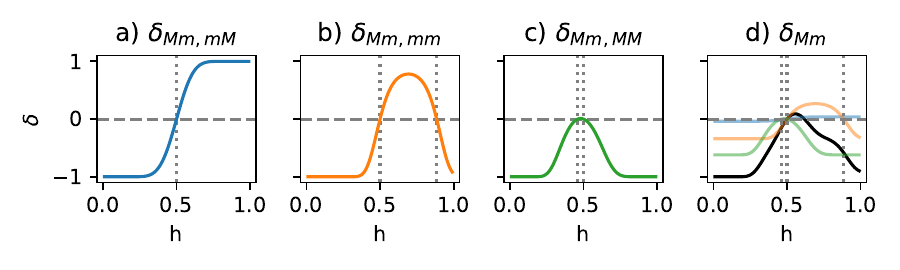}
\caption{\textbf{Understanding the trend of total disparity}. Panels \textbf{a}-\textbf{c} show the stochastic difference between the in-degree distribution of group $(M,m)$ and the rest in the system of Fig.~\ref{fig:examples_2D_inequality}\textbf{b}. Panel \textbf{d} shows the total disparity experienced by group $(M,m)$, which is a linear combination of the stochastic differences. The stochastic differences are shown as lighter color lines in panel \textbf{d} weighted by their corresponding coefficient according to Eq.~\eqref{eq:SI_ineq_linear_comb}: $\frac{F_{\boldsymbol{\sigma}}}{1-F_{Mm}}, \boldsymbol{\sigma}=\{mM,mm,MM\}$. Vertical lines mark the points where the in-degree difference changes sign according to Eqs.~\eqref{eq:SI_indegree_diffs}.}
\label{fig:ineq_decomp_example}
\end{figure}

In Fig.~\ref{fig:ineq_decomp_example} we show the $\delta_{\mathbf{r}, \mathbf{s}}$ comparing multidimensional group $(M,m)$ to each of the others in the system shown in Fig.~\ref{fig:examples_2D_inequality}\textbf{b}. We can appreciate how the stochastic differences $\delta_{\mathbf{r}, \mathbf{s}}$ present much simpler trends compared to total disparities $\delta_{\mathbf{r}}$, as the latter are linear combinations of the former. Stochastic differences are in turn determined by the expected in-degree of the groups, which depends on the population distribution and preference parameters:

\begin{equation}
    \lambda_\mathbf{r} = N \sum_{\mathbf{s}} F_\mathbf{s} H_{\mathbf{s},\mathbf{r}}
\end{equation}

In the 2D system with 2 categories per dimension studied in the paper, we set the one-dimensional preferences to:

\begin{equation}
    \begin{NiceMatrixBlock}[auto-columns-width]
    h^1_{r^1,s^1} = h^2_{r^2,s^2}= 
        \begin{bNiceMatrix}
            \mathsf{h} & 1-\mathsf{h} \\
           1-\mathsf{h} & \mathsf{h}
        \end{bNiceMatrix}
    \end{NiceMatrixBlock}
\end{equation}

To write the expression for $\lambda_\mathbf{r}$ for a focal group $(r^1,r^2)$  in this system, let us call $(r^1_C,r^2_C)$ the multidimensional group whose categories are the complementary of $(r^1,r^2)$, and let us call $(r^1_C,r^2)$ and $(r^1,r^2_C)$ the multidimensional group with one of the categories being the same and the other the complementary. With these considerations, we have:

\begin{equation}
    \lambda_\mathbf{r} = N [ F_{(r^1,r^2)} \mathsf{h}^2 + F_{(r^1_C,r^2_C)} (1-\mathsf{h})^2 + (F_{(r^1_C,r^2)} + F_{(r^1,r^2_C)}) \mathsf{h}(1-\mathsf{h})]
    \label{eq:SI_1D_indegre}
\end{equation}

This result shows that the expected in-group degree of multidimensional groups is a quadratic function of $\mathsf{h}$. Therefore, the difference of the expected in-group degree of two groups will also be a quadratic function of $\mathsf{h}$, changing sign at most in two points. According to Eq.~\eqref{eq:SI_multidim_1v1_ineq}, a multidimensional group $\mathbf{r}$ will be advantaged over a group $\mathbf{s}$  Iff. $\lambda_\mathbf{r}>\lambda_\mathbf{s}$. It is therefore of interest to compute for what values of $\mathsf{h}$ does the difference of expected in-degrees of two groups change sign. Using Eq.~\eqref{eq:SI_1D_indegre} and doing some simple algebra, we find:

\begin{equation}
\begin{aligned}
    \lambda_{(r^1,r^2)} - \lambda_{(r^1_C,r^2_C)} = N (F_{(r^1,r^2)}- F_{(r^1_C,r^2_C)})(2\mathsf{h}-1) =
    0 \quad \Leftrightarrow\quad \mathsf{h} = \frac{1}{2} \\ \\
    \lambda_{(r^1,r^2)} - \lambda_{(r^1_C,r^2)} = 
    N [F_{(r^1,r^2_C)}-F_{(r^1_C,r^2_C)}+
    \mathsf{h}(2 F_{(r^1,r^2)} + 2 F_{(r^1_C,r^2_C)} - 1)
    ]
    (2\mathsf{h}-1) =
    0 
    \\
    \Leftrightarrow\quad \mathsf{h}=\frac{1}{2}, \mathsf{h}=\frac{F_{(r^1_C,r^2_C) }- F_{(r^1,r^2_C)}}{2 F_{(r^1,r^2)} + 2 F_{(r^1_C,r^2_C)} - 1}\\ \\
    \lambda_{(r^1,r^2)} - \lambda_{(r^1,r^2_C)} = 
    N [F_{(r^1_C,r^2)}-F_{(r^1_C,r^2_C)}+
    \mathsf{h}(2 F_{(r^1,r^2)} + 2 F_{(r^1_C,r^2_C)} - 1)
    ]
    (2\mathsf{h}-1) =
    0 
    \\
    \Leftrightarrow\quad \mathsf{h}=\frac{1}{2}, \mathsf{h}=\frac{F_{(r^1_C,r^2_C) }- F_{(r^1_C,r^2)}}{2 F_{(r^1,r^2)} + 2 F_{(r^1_C,r^2_C)} - 1}
\end{aligned}
\label{eq:SI_indegree_diffs}
\end{equation}

These equations show that $\lambda_\mathbf{r}-\lambda_\mathbf{s}$ in the considered system always changes sign at $\mathsf{h}=0.5$. If the groups are completely complimentary (such as $mm$ and $MM$), that is the only point where the in-degree difference changes sign. If the groups differ in only one of the categories (such as $mm$ and $mM$), the in-degree difference may also change sign in another point as long as the second solution falls within the interval $\mathsf{h}\in[0,1]$.

If the attributes are uncorrelated, so that in the 2D system $F_{(r^1,r^2)}=f^1_{r^1} f^2_{r^2}$, the relationships in Eq.~\eqref{eq:SI_indegree_diffs} simplify to:

\begin{equation}
\begin{aligned}
    \lambda_{(r^1,r^2)} - \lambda_{(r^1_C,r^2_C)} = N (f^1_{r^1} + f^2_{r^2} - 1)(2\mathsf{h}-1) =
    0 \quad \Leftrightarrow\quad \mathsf{h} = \frac{1}{2} \\ \\
    \lambda_{(r^1,r^2)} - \lambda_{(r^1_C,r^2)} = 
    N
    (2\mathsf{h}-1) 
    (2f^1_{r^1}-1)
    [1-f^2_{r^2}-h(1-2f^2_{r^2})]
    =
    0 
    \\
    \Leftrightarrow\quad \mathsf{h}=\frac{1}{2}, 
    \mathsf{h}=\frac{1-f^1_{r^1}}{1-2f^1_{r^1}}\\ \\
    \lambda_{(r^1,r^2)} - \lambda_{(r^1,r^2_C)} = 
    N 
    (2\mathsf{h}-1) 
    (2f^2_{r^2}-1)
    [1-f^1_{r^1}-h(1-2f^1_{r^1})]=
    0 
    \\
    \Leftrightarrow\quad \mathsf{h}=\frac{1}{2}, 
    \mathsf{h}=\frac{1-f^2_{r^2}}{1-2f^2_{r^2}}
\end{aligned}
\label{eq:SI_indegree_diffs_uncorr}
\end{equation}

Notice that the second solution of the last two equations is always $\frac{1-f^d_{r^d}}{1-2f^d_{r^d}}>1$, so when attributes are uncorrelated there is only one change of sign at $\mathsf{h}=0.5$. Furthermore, the last factor of both equations is always $[1-f^d_{r^d}-h(1-2f^d_{r^d})]>0$, since $0\leq\mathsf{h}\leq1$ and   $(1-f^d_{r^d}) >(1-2f^d_{r^d} ) \forall f^d_{r^d} \in \mathbb{R}$ .  Therefore, the sign of the in-degree differences is controlled by the factor $(2\mathsf{h}-1)$, which depends on whether the system is homophilic $(\mathsf{h}>0.5)$ or heterophilic $(\mathsf{h}<0.5)$, and the factor $(2f^d_{r^d}-1)$, which is positive when $f^d_{r^d}>0.5$; that is, if the one-dimensional group $r^d$ is the majority in dimension $d$, and negative when $f^d_{r^d}<0.5$.

This result implies that when attributes are uncorrelated in the considered system, the advantages and disadvantages of multidimensional groups (and therefore, also of one-dimensional groups), are structured qualitatively in the same way as in binary one-dimensional systems, with smaller groups always being advantaged in the heterophilic regime and larger groups in the homophilic regime. Therefore, in this system, we can only have emergent intersectionality if attributes are correlated. If we label the one-dimensional groups so that $f^1_m \leq f^2_m \leq f^2_M \leq f^1_M$, with uncorrelated attributes we would always find $\delta_{MM}\geq\delta_{Mm}\geq\delta_{mM}\geq\delta_{mm}$ for $\mathsf{h}>0.5$ and the opposite for $\mathsf{h}<0.5$, like in Fig.~\ref{fig:examples_2D_inequality}\textbf{a}. More complex behaviors (as in panels \textbf{b}-\textbf{f} of Fig.~\ref{fig:examples_2D_inequality}) only emerge in this system when there is attribute correlation.

\section{Complete inequality maps for a 2D system with symmetric homophily}
\label{sec:SI_inequality_maps}
\begin{figure}[h!]
\centering
\includegraphics[width=1.0\textwidth]{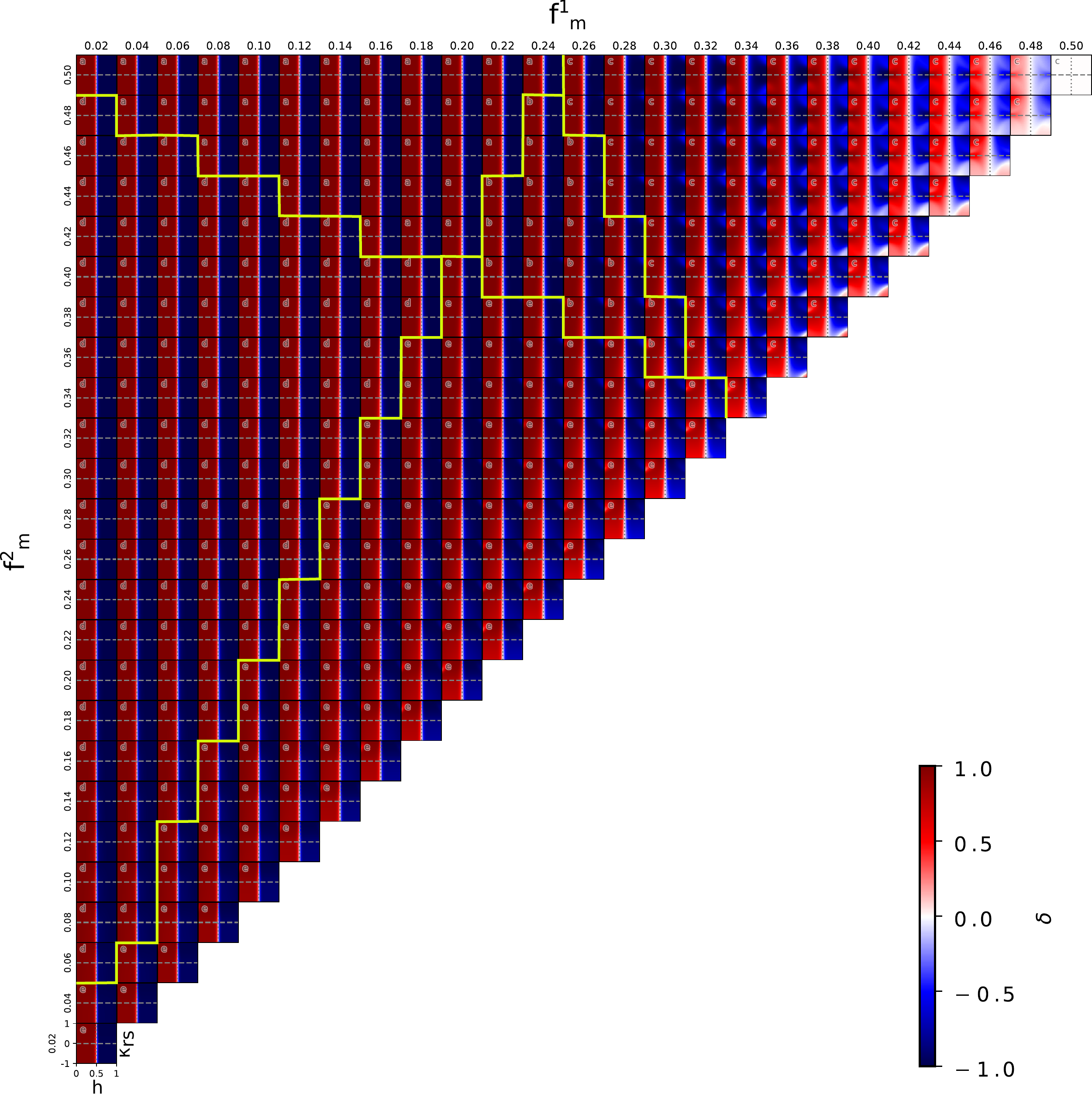}
\caption{\textbf{Inequality maps for minority of dimension 1}.}
\label{fig:ineq_map_m1}
\end{figure}

\begin{figure}[h!]
\centering
\includegraphics[width=1.0\textwidth]{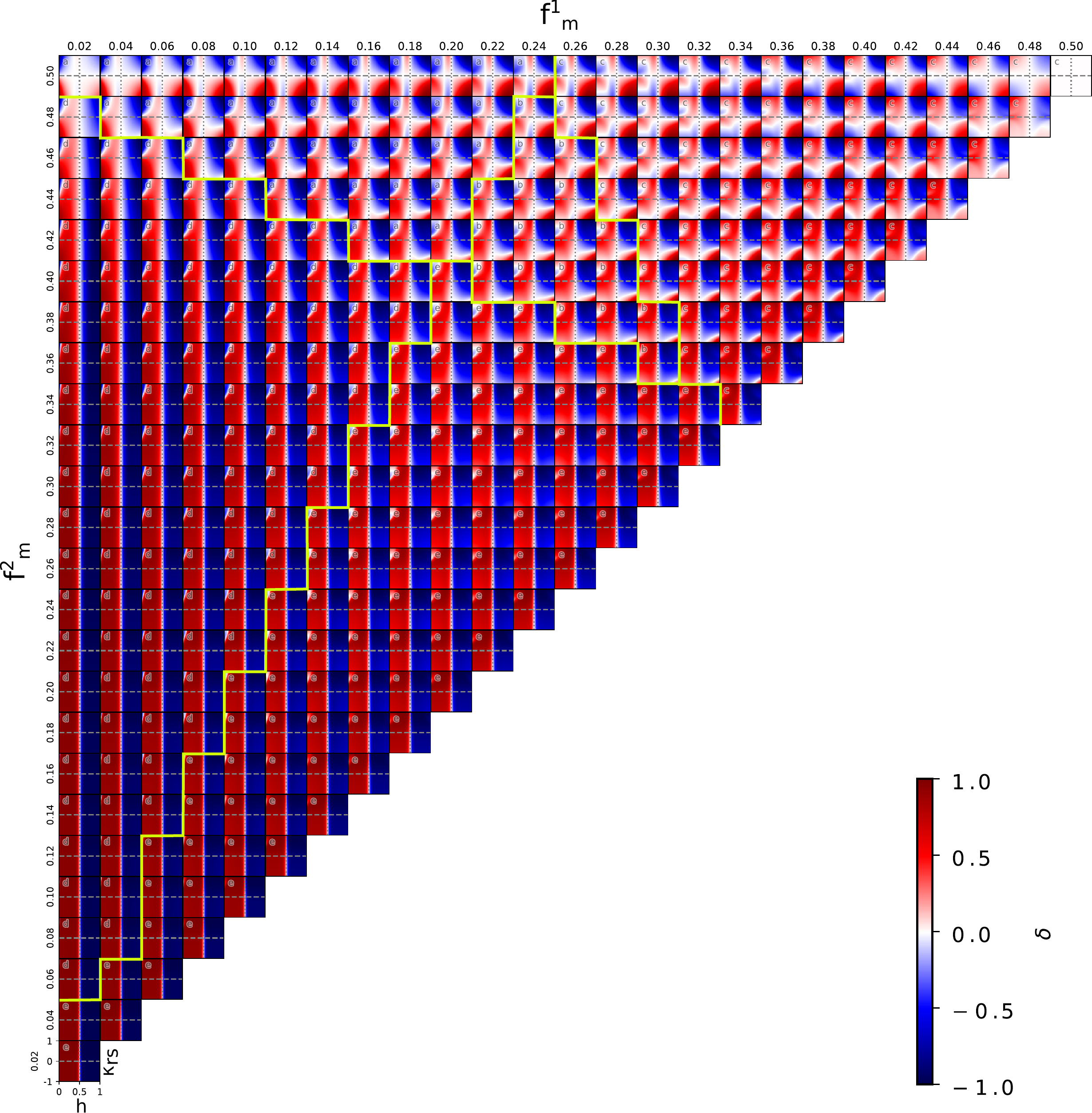}
\caption{\textbf{Inequality maps for minority of dimension 2}.}
\label{fig:ineq_map_m2}
\end{figure}

\begin{figure}[h!]
\centering
\includegraphics[width=1.0\textwidth]{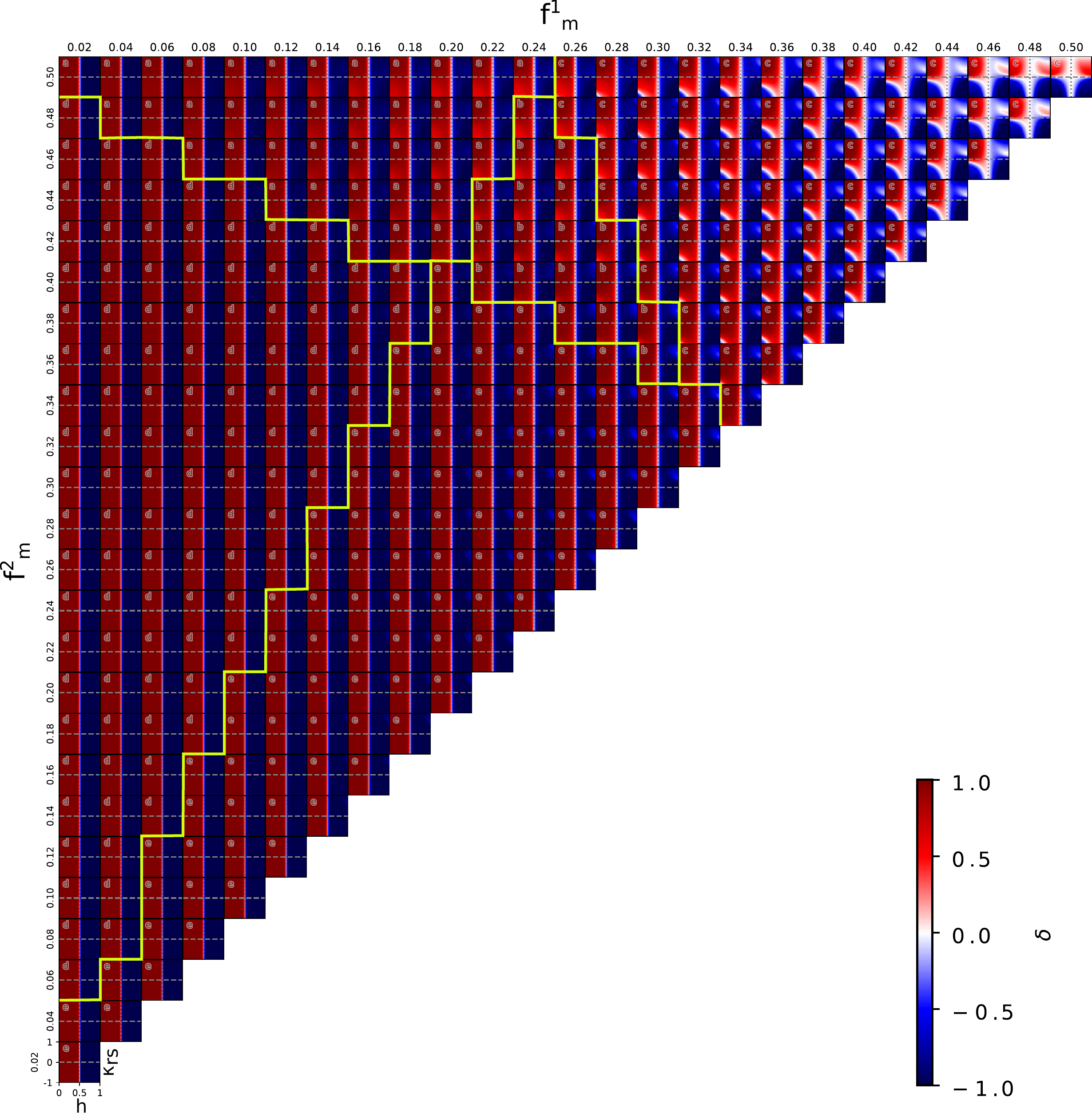}
\caption{\textbf{Inequality maps for multidimensional group $mm$}.}
\label{fig:ineq_map_00}
\end{figure}

\begin{figure}[h!]
\centering
\includegraphics[width=1.0\textwidth]{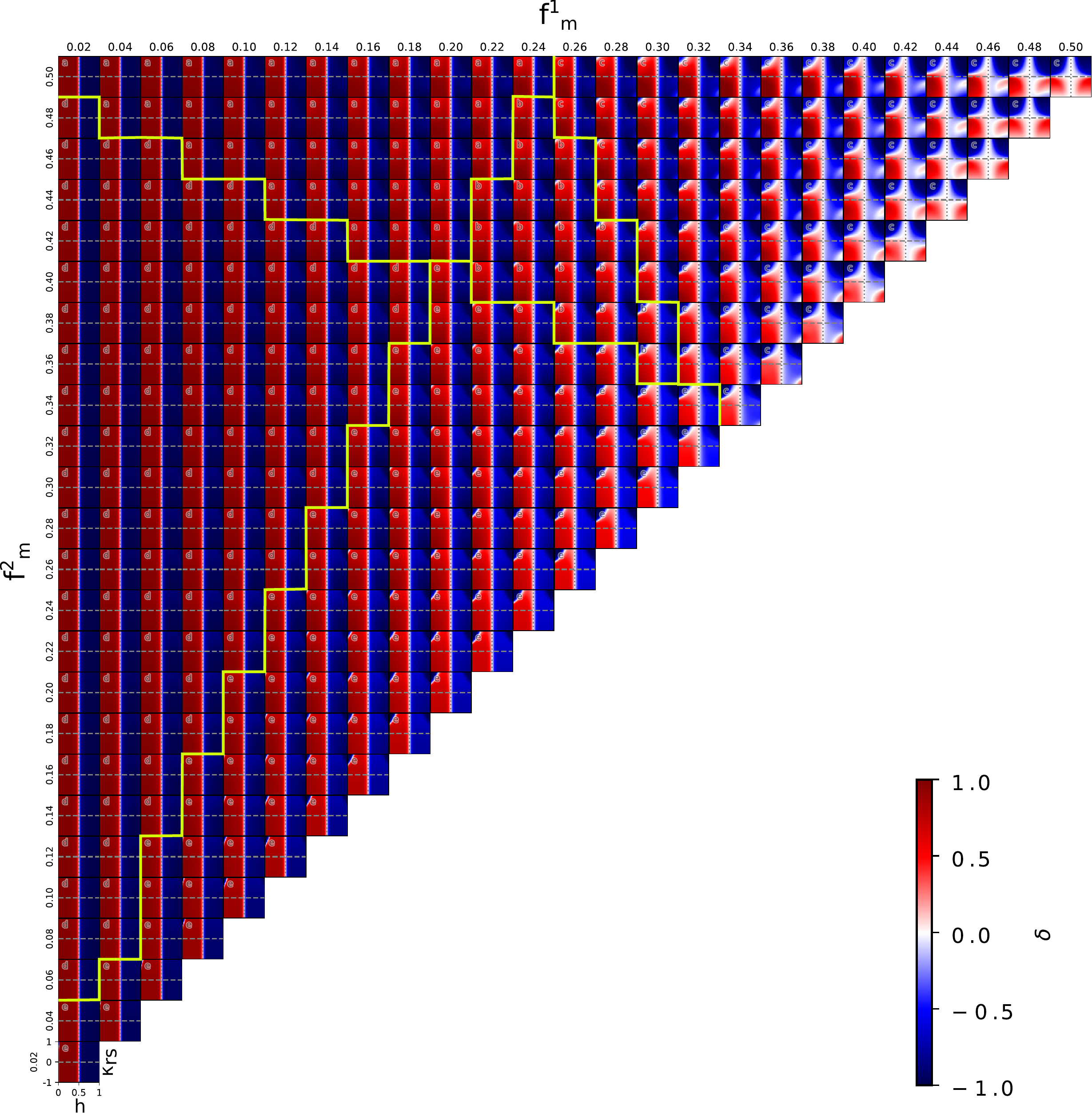}
\caption{\textbf{Inequality maps for multidimensional group $mM$}.}
\label{fig:ineq_map_01}
\end{figure}

\begin{figure}[h!]
\centering
\includegraphics[width=1.0\textwidth]{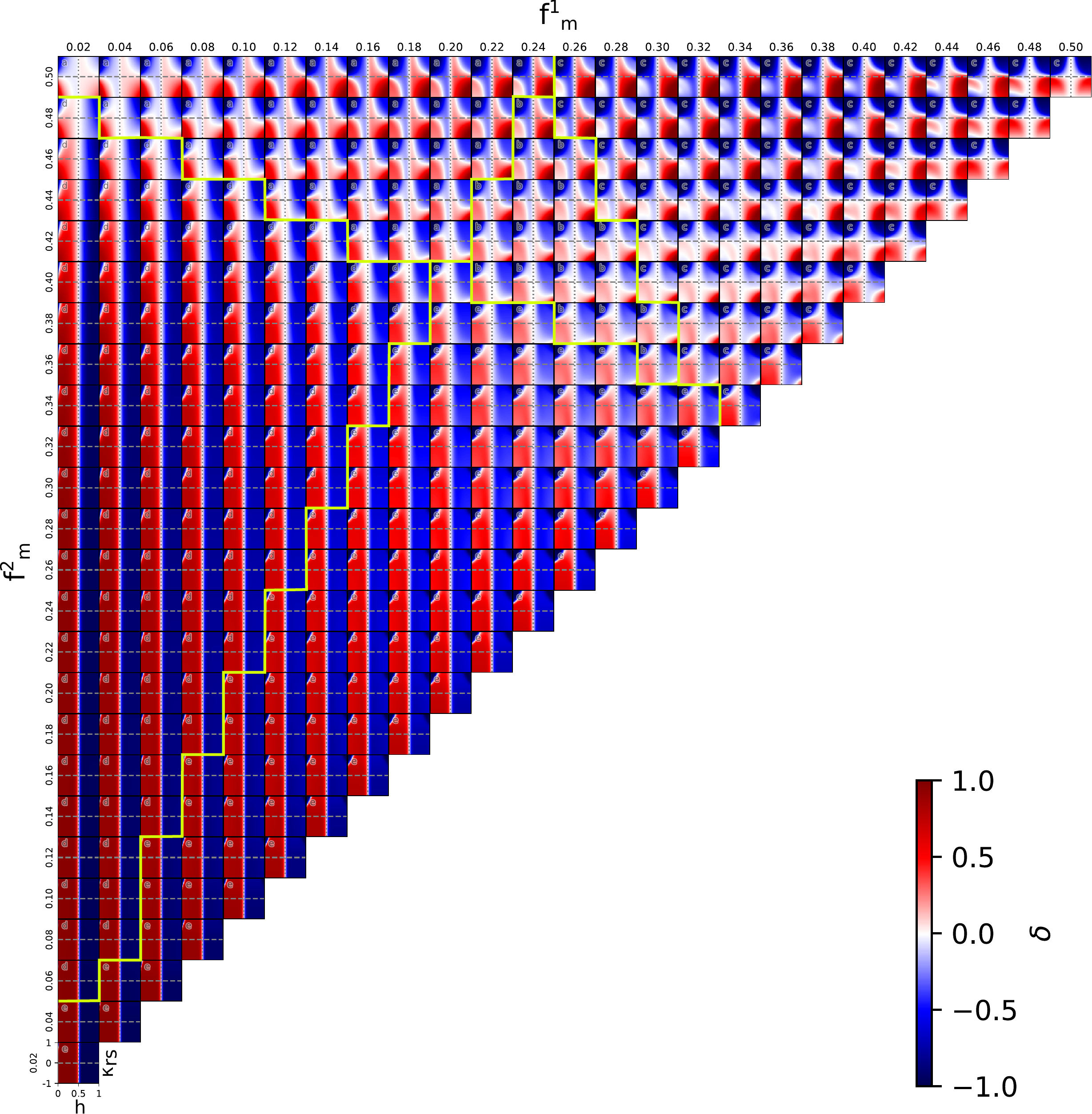}
\caption{\textbf{Inequality maps for multidimensional group $Mm$}.}
\label{fig:ineq_map_10}
\end{figure}

\begin{figure}[h!]
\centering
\includegraphics[width=1.0\textwidth]{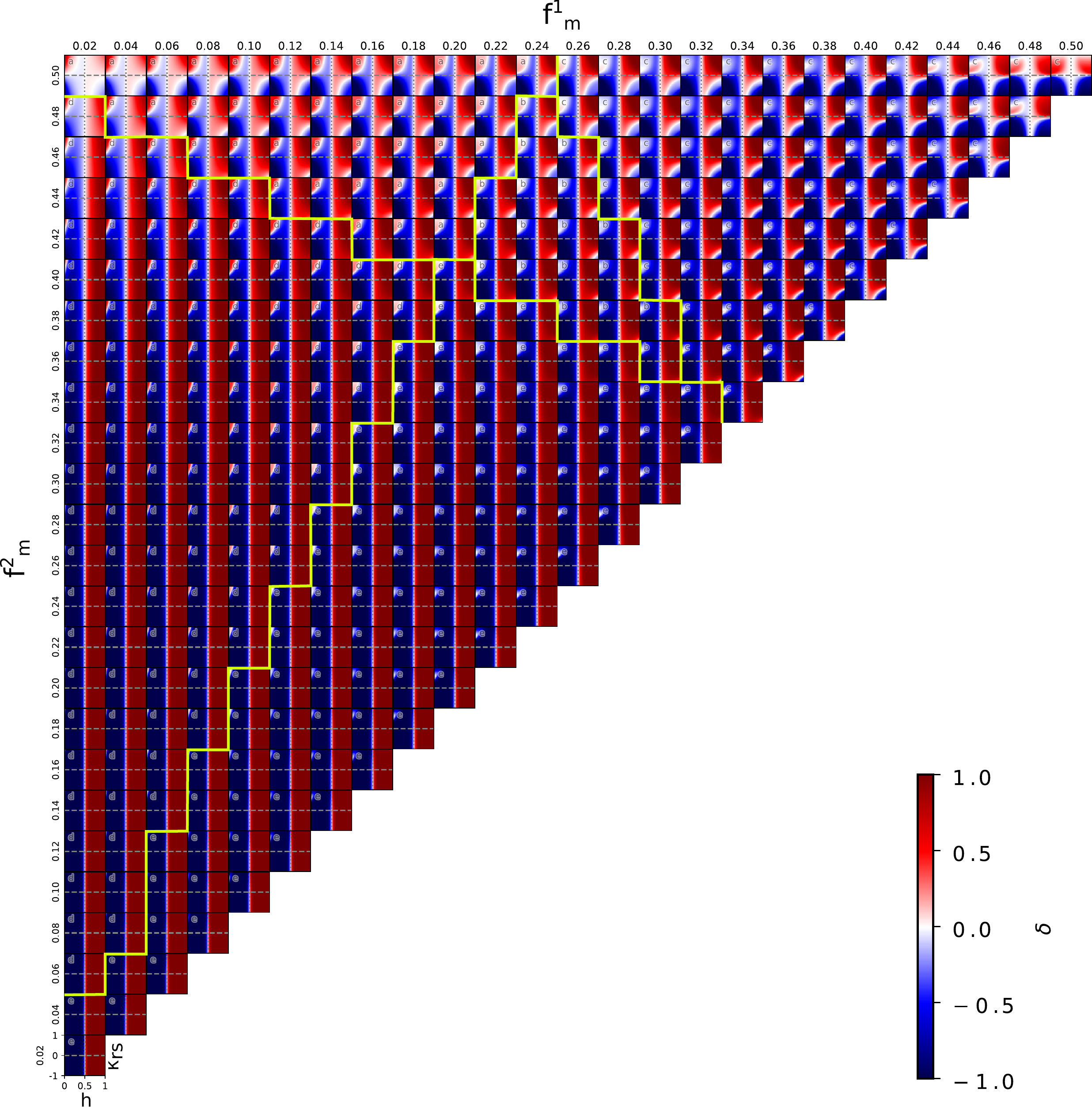}
\caption{\textbf{Inequality maps for multidimensional group $MM$}.}
\label{fig:ineq_map_11}
\end{figure}

In this section, we present a comprehensive characterization of inequalities $\delta$ in a 2D system with latent preferences given by Eq. \eqref{eq:1dpref_tune_h}. We systematically tune the 4 model parameters: $f^1_m, f^2_m, \kappa, \mathsf{h}$. Inequalities are shown as heatmaps in Figs. \ref{fig:ineq_map_m1}-\ref{fig:ineq_map_11}. Each mini panel corresponds to a system with different minority fractions $(f^1_m,f^2_m)$. The panels show heatmaps where the $x$ axis is the homophily parameter $\mathsf{h}$, the $y$ axis is the rescaled correlation parameter $\kappa_{rs}$, and the color is $\delta$, with blue and red representing negative and positive values respectively. The yellow borders between the mini panels divide the regions of the $(f^1_m,f^2_m)$ space defined in Fig. \ref{fig:effect_of_population_distribution}, and each mini panel is labeled with the region it belongs to.

\end{document}